\documentclass[final]{elsart}
\usepackage[dvips]{graphicx}
\usepackage{amssymb}
\usepackage{amsmath}
\usepackage{subfigure} 
\usepackage[usenames]{color}
\usepackage{color}

\begin{document}

\begin{frontmatter}

\title{A new composition-sensitive parameter for Ultra-High Energy Cosmic Rays}

\author[ALCALA]{G. Ros},
\author[UNAM]{A. D. Supanitsky},
\author[UNAM]{G. A. Medina-Tanco},
\author[ALCALA]{L. del Peral},
\author[UNAM]{J. C. D'Olivo}, and
\author[ALCALA]{M. D. Rodr\'iguez Fr\'ias}
\address[ALCALA]{Space and Astroparticle Group, Dpto. F\'isica, Universidad de Alcal\'a 
Ctra. Madrid-Barcelona km. 33. Alcal\'a de Henares, E-28871 (Spain).} 
\address[UNAM]{Instituto de Ciencias Nucleares, UNAM, Circuito Exterior S/N, Ciudad Universitaria,
M\'exico D. F. 04510 (Mexico).}

\begin{abstract}

A new family of parameters intended for composition studies in cosmic ray surface array detectors is proposed. The application of this technique to different array layout designs has been analyzed. The parameters make exclusive use of surface data combining the information from the total signal at each triggered detector and the array geometry. They are sensitive to the combined effects of the different muon and electromagnetic components on the lateral distribution function of proton and iron initiated showers at any given primary energy. Analytical and numerical studies have been performed in order to assess the reliability, stability and optimization of these parameters. Experimental uncertainties, the underestimation of the muon component in the shower simulation codes, intrinsic fluctuations and reconstruction errors are considered and discussed in a quantitative way. The potential discrimination power of these parameters, under realistic experimental conditions, is compared on a simplified, albeit quantitative way, with that expected from other surface and fluorescence estimators.

\end{abstract}


\end{frontmatter}

\section{Introduction} \label{Introduction}

Ultra-high energy cosmic rays (UHECR) are studied by the detection of the extensive air showers
(EAS) that they produce in the atmosphere. At present, there are two main observation techniques. One, the so called fluorescence technique, employs telescopes to collect the fluorescence light emitted by atmospheric Nitrogen molecules after being excited by the charged particles of the cascade. The other one, the surface technique, is based on the analysis of a discrete sampling of the shower front at the ground level, performed by surface detectors. UHECRs produce few observables. They are, basically, the arrival direction, the energy and some statistical hint about the identity of the primary particle. Of these three pieces of information, the geometrical one is the most reliable for both detection techniques \cite{AugerAngularError, AGASAAngular, HiResAngular, HPAngular, TAStatus}.
The energy of the shower can be inferred with an accuracy of around 20\% in the case of stereo fluorescence reconstruction \cite{HiResSpectrumStereo}. In the case of hybrid detectors, this accuracy can be transferred to surface arrays using as calibrators events which are observed simultaneously by both techniques. The Auger Observatory was the pioneer of this strategy  \cite{AugerHybridSpectrumICRC09} which is likely to become a standard for accuracy in the field \cite{TAStatus, AugerNorth}.  The need for such cross calibration already 
highlights the existence of as yet either unidentified problems with our understanding of the physics involved in shower generation and development or inconsistencies in our implementation of those physical processes into the available shower simulation codes. Most of these problems certainly have their roots in the uncertainties associated with the extrapolation of cross sections, multiplicities and inelasticities from accelerator measurements at much lower energies, required to treat the first hadronic interactions experienced by the incoming cosmic ray in the upper layers of the atmosphere. Given the indirect nature of the detection of cosmic rays at the highest energies, those uncertainties permeate, to a larger or lesser extent, all measurements done afterwords. In particular, they have their strongest manifestation on the interpretation and reliability of mass composition tracers, since variations in cross section or inelasticity can easily be misinterpreted as changes in baryonic composition (see \cite{Knapp-HM} and \cite{Haungs-HM} for a thorough discussion on the relevance of hadronic interaction models to shower development).

Each observation technique has specific composition indicators. The fluorescence technique is, at present, the most reliable one for composition studies since the longitudinal development of the charged component of the atmospheric shower is a relatively straightforward observable. Differences in composition, manifest themselves through differences in the cross section for interactions with atmospheric nuclei. These, in turn, are mapped as different average depths of maximum development of the electromagnetic component in the atmosphere ($X_{max}$) for different nuclei and as different statistical dispersions for the $X_{max}$ distributions of different nuclei ($\Delta X_{max}$). If, for example, proton and Iron primaries are compared, the smaller cross section of the former will produce larger $X_{max}$ and $\Delta X_{max}$ than the latter \cite{AugerXmax2010, HiResElongationRate}. However, unforeseen changes in cross section as a function of energy can affect these parameters in much the same way as true changes in composition would \cite{Ulrich}.

Surface detectors, on the other hand, sample the lateral distribution function (LDF) of the EAS at discrete 
points while they traverse the ground level. Beyond a few tens of meters from the shower axis, the particle content of the shower at ground level is dominated by just two components, electromagnetic (i.e., electrons, positrons and photons) and muonic. These two sets of particles propagate differently through the atmosphere: the electromagnetic component propagates diffusively, while the muons do so radially from the last hadronic interaction region, which produces their parent mesons. Therefore, in a simplified way, the shower front can be thought of as the composition of two fronts, a muonic one, which arrives first and is temporally thin and an electromagnetic disk, more extended in time that follows the muon front. Furthermore, the muon shower front has a much better defined and larger curvature radius than the electromagnetic front. One of the practical effects of these differences is that information about the relative intensity of both components inside the shower and, therefore, of the identity of the primary nuclei, is encoded simultaneously in a non-trivial way in several properties of the shower front at ground level: the slope of the lateral distribution function of particles, the radius of curvature and its time structure. Thus, several parameters have been proposed to extract composition information from the surface measurements of EAS. The discrimination power of each parameter depends on the type of detector selected, water Cherenkov tanks (e.g., Haverah Park, Auger) or scintillators (e.g., Volcano Ranch, AGASA, Telescope Array). The former are sensitive to both the electromagnetic and muonic components (although the number of muons is much smaller than the electromagnetic 
particles, they generate a large amount of Cherenkov photons in the tanks) while the latter are sensitive to the charged particles of the cascade which are dominated by electrons and positrons. Scintillators are also shielded or buried underground to measured the muonic component of the showers. The most useful parameters so far have been the slope of the LDF \cite{HP-Composition, VR-Composition}, the curvature of the shower front, the number of muons in the shower \cite{Supa1, Supa2}, several indicators of the time structure of the shower like the rise time and other related parameters \cite{HP-Composition, AugerCompositionSDICRC07, AugerCompositionSDICRC09}, and the azimuthal asymmetries in the signals \cite{Garcia-Pinto}. 

In general terms, fluorescence composition indicators are regarded as easier to observe and interpret, as well as less prone to systematic errors than surface parameters do. However, fluorescence detectors are constrained by a duty cycle of approximately 10\%, while well operated, stable surface detectors can reach duty cycles near 100\%. This factor alone, which makes the statistics per unit time of surface arrays an order of magnitude larger than that of fluorescence detectors, gives a great attractive to the search for reliable surface composition parameters.

In the present work, we propose a new surface parameter which, we argue, for the same integration time can deliver better discrimination power than $X_{max}$. The proposed parameter is defined as:
\begin{equation}\label{Sb}
S_b = \sum_{i=1}^{N} \left[ S_{i} \times \left(\frac{r_i}{r_0}\right)^b \right]
\end{equation}
where the sum extends over all the triggered stations \emph{N}, $r_0=1000$ m is a reference distance, $S_{i}$ is the signal measured at the $i$th station and $r_{i}$ is the distance of this station to the shower axis in meters. $S_b$ is sensitive to the combined effects of the different muon and electromagnetic components on the lateral distribution function of proton and Iron initiated showers.

As a case study, we analyze hereafter the behavior and main properties of $S_b$ by assuming general arrays of water Cherenkov detectors. The results, however, should be still qualitatively valid for any other detector which produces comparable signals for the muonic and electromagnetic components of the showers. Therefore, the signal and $S_b$ are measured in VEM (Vertical Equivalent Muons, as in Haverah Park and Auger experiments). In this case, we will demonstrate later that the primary identity discrimination power goes through a maximum around $b=3$. The analysis has been performed considering different array geometries and varying the detectors separation, showing the general applicability of the parameter. $S_b$ could also be applied to scintillator arrays as the future Telescope Array experiment \cite{TAStatus}.

The paper is organized as follows: Section 2 shows an analytical discussion of the properties and stability of the new parameter.
Section 3 shows in some respects an equivalent numerical study performed with simulations taking into account the effects 
of reconstruction and experimental uncertainties. In Section 4 we perform a realistic comparative study among the stability 
and reliability of the inferred composition by using $S_3$, $X_{max}$ and the rise time at 1000 m from core. Conclusions 
are presented in Section 5.

\section{Analytical study} \label{analytical_analysis}

\begin{figure}[!hb]
\begin{center}
\includegraphics[width=6.8cm]{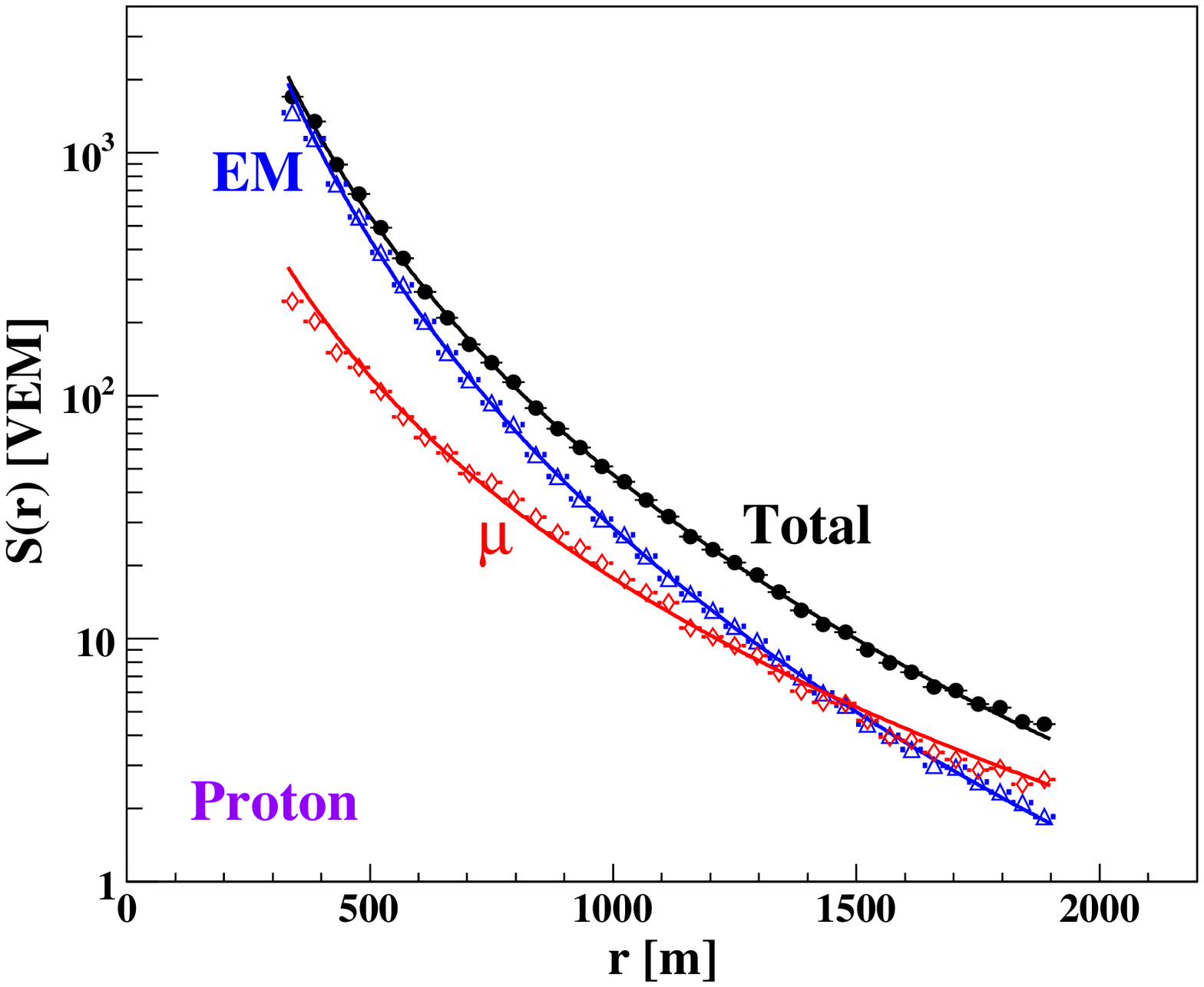}
\includegraphics[width=6.8cm]{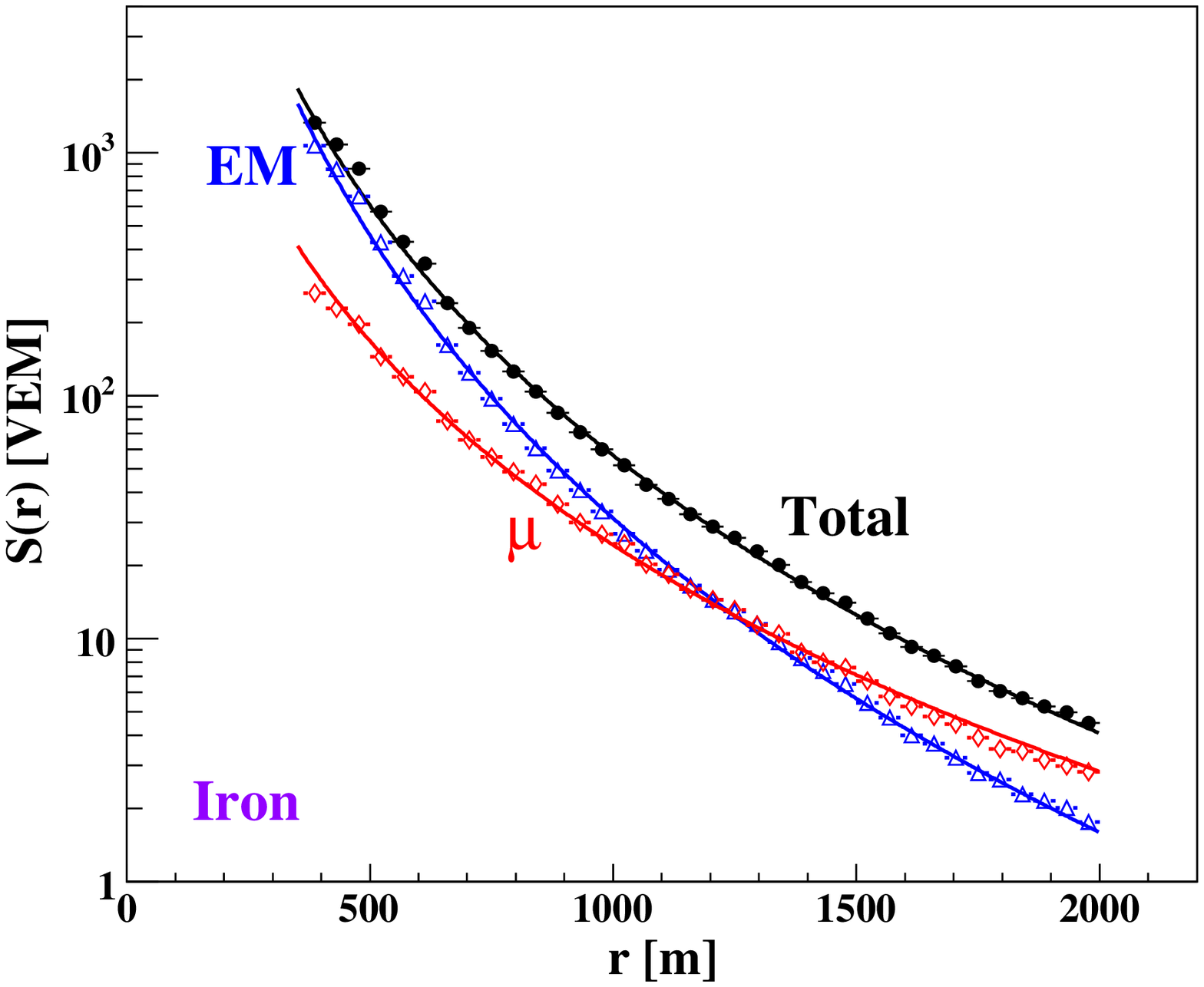}
\caption{Lateral distribution functions of the muon, electromagnetic and total signal in the Cherenkov detectors for 
simulated protons (left) and iron nuclei (right) of $1 \leq \sec\theta \leq 1.2$ and $19 \leq \log(E/\textrm{eV}) \leq 19.1$. 
The hadronic interaction model used to generate the showers is QGSJET-II. The solid lines correspond to the fits with a 
NKG-like function (see Eq. (\ref{NKGldf})). \label{FitLdfs}}
\end{center}
\end{figure}

The parameter $S_b$ for a given event is constructed from the total signal in each triggered Cherenkov detector. Therefore, it depends on the normalization and shape of the lateral distribution function of the total signal. Close to the impact point of the shower, the signal is dominated by the electromagnetic particles (photons, electrons and positrons) whereas at larger distances it is dominated by muons. Fig. \ref{FitLdfs} shows the muon, electromagnetic and total signal in the Cherenkov detectors as a function of the distance to the shower axis for protons and Iron nuclei. The zenith angle of the simulated events considered is such that $1 \leq \sec\theta \leq 1.2$ and the primary energy $19 \leq \log(E/\textrm{eV}) \leq 19.1$ (see section \ref{Numerical_Analysis} for details about the simulations). The hadronic model considered is QGSJET-II. Fig. \ref{FitLdfs} also shows the fits of the LDF of each component with a NKG-like function \cite{NKG,ICRCldf:05,TancoICRC05},
\begin{equation} 
\label{NKGldf}
S(r)=S_0 \left( \frac{r}{r_0} \right)^\beta \left( \frac{r+r_s}{r_0+r_s} \right)^\beta, 
\end{equation}
where $r_s=700$ m and $r_0=1000$ m, and $S_0$ and $\beta$ are free fit parameters.
If we consider proton and Iron primaries, the discrimination power of a mass sensitive parameter $q$, like $S_b$, can be 
estimated by using the so-called merit factor defined as
\begin{equation}  
\label{Eta}
\eta = \frac{E[q_{fe}]-E[q_{pr}]}{\sqrt{Var[q_{fe}]+Var[q_{pr}]}},
\end{equation}
where $E[q_A]$ and $Var[q_A]$ are the mean value and the variance, respectively, of the distribution function of the
parameter $q_A$ with $A=pr, fe$. Note that an alternative definition for the merit factor makes use of the median 
instead of the mean value and, instead of the variance, $\sigma_{68}^2[q] \equiv [(q_{84}-q_{16})/2]^2$, where 
$q_{84}$ and $q_{16}$ are the quantiles corresponding to 84\% and 16\% of probability, respectively. Now we use 
the definition as it is in Eq. (\ref{Eta}) to make possible the analytical analysis.

\subsection{Optimization in water Cherenkov detectors arrays}

Assuming that the fluctuations of the total signal in 
a water Cherenkov detector are Gaussian, the distribution function for a given configuration of triggered stations 
is given by,
\begin{equation}
\label{Dist}
P(s_1, \ldots ,s_N; r_1, \ldots ,r_N) = \frac{f(r_1,\ldots,r_N)}{(2\pi)^{N/2} \prod^{N}_{i=1}\sigma[S(r_i)]}%
\exp\left[-\sum^{N}_{i=1} \frac{(s_i-S(r_i))^2}{2\ \sigma^2[S(r_i)]} \right],
\end{equation}
where $r_i$ is the distance to the shower axis of the $i$th station (the first station, $r_1$, is the closest one), 
$S(r_i)$ is the average LDF evaluated at $r_i$, $\sigma[S(r_i)]=(1+\varepsilon)\ [S(r_i)/\textrm{VEM}]^{1/2}$ VEM \cite{ HP-Composition, ICRCldf:05} and $f(r_1,\ldots,r_N)$ is the distribution function of the distance of the different stations to the shower axis. We have checked that for values of $\varepsilon$ of at least 5\% there is no change in the results. Note that just two of the random variables $\{r_1, \ldots, r_N\}$ are independent, for instance, choosing $r_1$ and $r_2$ (the first and second closest stations to shower axis) as the independent ones, we can write $f(r_1,\ldots,r_N)=f_{1,2}(r_1,r_2)\delta(r_3-r_3(r_1,r_2))\ldots\delta(r_N-r_N(r_1,r_2))$, where $\delta(x)$ is the Dirac delta function.

From Eqs. (\ref{Sb}) and (\ref{Dist}) we obtain the expressions for the mean value and the variance of $S_b$,
\begin{eqnarray}
\label{MVSb}
E[S_b] &=& \sum_{i=1}^N E\left[ S(r_i) \left( \frac{r_i}{r_0}  \right)^b \right], \\
Var[S_b] &=& (1+\varepsilon)^2\ \sum_{i=1}^N E\left[ S(r_i) \left( \frac{r_i}{r_0}  \right)^{2 b} \right]+ \nonumber \\%
&&\sum_{i=1}^N \sum_{j=1}^N cov\left[ S(r_i) \left( \frac{r_i}{r_0}  \right)^b, S(r_j) \left( \frac{r_j}{r_0} \right)^b \right],
\end{eqnarray}
where the variables $s_i$ have already been integrated and,
\begin{eqnarray}
\label{Int}
E\left[ S(r_i) \left( \frac{r_i}{r_0}  \right)^x \right] &=& \int dr_i\ S(r_i) \left( \frac{r_i}{r_0} \right)^x f_i(r_i), \\
cov\left[ S(r_i) \left( \frac{r_i}{r_0}  \right)^b\!, S(r_j) \left( \frac{r_j}{r_0} \right)^b \right] &=&%
\int dr_i\ dr_j\ S(r_i) \left( \frac{r_i}{r_0} \right)^b S(r_j) \left( \frac{r_j}{r_0} \right)^b \times    \nonumber \\
&& f_{i,j}(r_i,r_j).
\end{eqnarray}
Here, $f_i(r_i)$ is the distribution function of the distance to the shower axis for the $i$th station and $f_{i,j}(r_i,r_j)$ 
is the distribution function of the distance to the shower axis of the $i$th and $j$th stations,
\begin{equation}
\label{Fr1r2}
f_{i,j}(r_i,r_j) = \int dr_1 \ldots dr_{i-1} dr_{i+1} \ldots dr_{j-1} dr_{j+1} \ldots dr_N\ f(r_1,\ldots,r_N).
\end{equation}
In order to simplify the expressions for the mean and variance of $S_b$, we perform the following approximations,
\begin{eqnarray}
\label{App}
E[g(r_i)] &\cong& g(E[r_i]), \\
cov[g(r_i), g(r_j)] &\cong& \left. \frac{dg}{dr} \right|_{E[r_i]} \left. \frac{dg}{dr}\right|_{E[r_j]} cov[r_i,r_j],
\end{eqnarray}
where $g(r)=S(r) (r/r_0)^b$. Thus, we finally get,
\begin{eqnarray}
\label{FinalFormulas}
E[S_b] &=& \sum_{i=1}^N S(E[r_i]) \left( \frac{E[r_i]}{r_0}  \right)^b  \\ 
Var[S_b] &=& (1+\varepsilon)^2\ \sum_{i=1}^N S(E[r_i]) \left( \frac{E[r_i]}{r_0}  \right)^{2 b} +  \nonumber \\
&&\sum_{i=1}^N \sum_{j=1}^N \frac{\partial}{\partial r_i} \left( S(r_i) \left. \left( \frac{r_i}{r_0} \right)^{b} \right) 
\right|_{E[r_i]} \frac{\partial}{\partial r_j} \left( S(r_j) \left. \left( \frac{r_j}{r_0} \right)^{b} \right) 
\right|_{E[r_j]} \times \nonumber \\
&&cov[r_i,r_j].
\end{eqnarray}

\begin{figure}[!bt]
\begin{center}
\includegraphics[width=12cm]{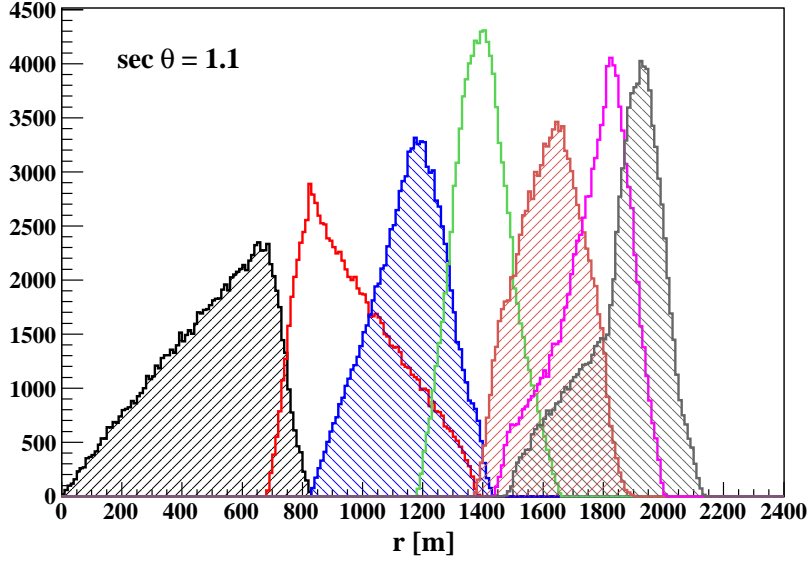}
\caption{Distance of the stations to shower axis for almost vertical showers in a triangular grid of 1.5 km of spacing. 
\label{MCStations}}
\end{center}
\end{figure}

We already have analytical expressions for the average LDFs of proton and Iron primaries obtained by fitting the simulated data 
(Fig. \ref{FitLdfs}). The other ingredients needed to calculate the mean value and the variance of $S_b$ are the mean values of 
the distance to the shower axis for the different stations and the covariance between all pairs of those random variables. We 
obtain these quantities from a simple Monte Carlo simulation: we uniformly distribute impact points in a triangular grid of 
1.5 km of spacing, like the Auger array, and then, for each event of zenith angle such that $\sec\theta=1.1$ and azimuthal 
angle uniformly distributed in $[0,2 \pi]$, we calculate the distance of each station to the shower axis. The result is shown 
in Fig. \ref{MCStations}. From these distributions $E[r_i]$ and $cov[r_i,r_j]$ are easily determined.

\begin{figure}[!bt]
\begin{center}
\includegraphics[width=10cm]{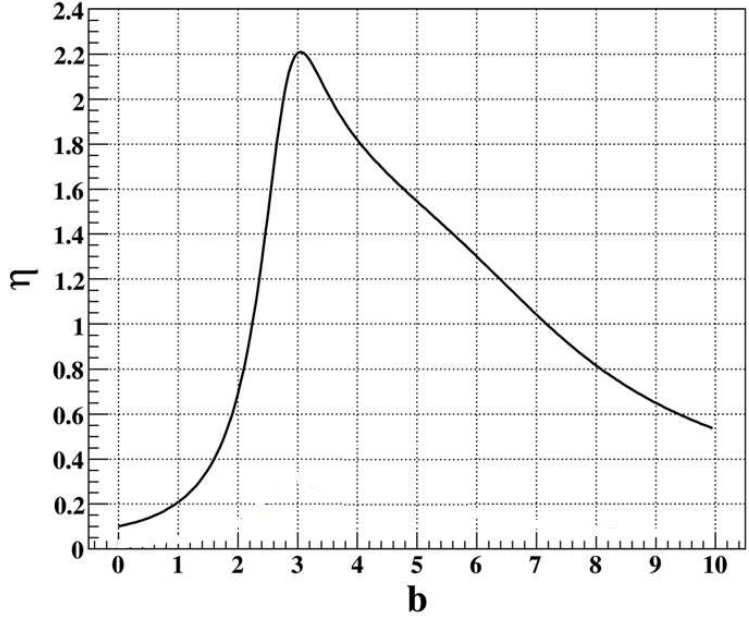}
\caption{$\eta$ as a function of $b$ for vertical showers ($1 \leq \sec\theta \leq 1.2$) and 
$19 \leq \log(E/eV) \leq 19.1$. $\eta$ reaches the maximum at $b\cong3$. \label{EtaVsB}}
\end{center}
\end{figure}

Finally, we could calculate the mean and the variance of $S_b$, and therefore, the merit factor. 
Fig. \ref{EtaVsB} shows the discrimination power $\eta$ as a function of $b$ obtained under the mentioned assumptions 
and simplifications. We see that $\eta$ reaches the maximum at $b\cong3$. 

\subsection{Modifying the slope of the LDF}

We also study the discrimination power of $S_b$ when the slope parameter $\beta$ (see Eq. (\ref{NKGldf})) is modified but keeping constant the 
integrated signal for distances larger than the Moliere radius, $r_M=80$ m. Thus, the inferred energy of the event by 
surface experiments would not be significantly affected. The modified LDF that fulfills this condition can be written as,
\begin{equation}
\label{LDFAC}
S(r, \beta)=\frac{N(r_M, r_0, r_s, \beta_0)}{N(r_M, r_0, r_s, \beta)}\ S_{\beta_0}(r), 
\end{equation}
where,
\begin{equation}
\label{Norm}
N(r_M, r_0, r_s, \beta) = \frac{r_s^{2+2\beta}}{r_0^\beta (r_s+r_0)^\beta}%
\ \textrm{Beta}(-r_s/r_M,-2 (1+\beta), 1+\beta),
\end{equation}
$\textrm{Beta}(z,a,b)=\int_0^z dt\ t^{a-1} (1-t)^{b-1}$ and $S_{\beta_0}(r)$ is the LDF of Eq. (\ref{NKGldf}) with 
the parameters $S_0$ and $\beta_0$ originally obtained from the fits in Fig. \ref{FitLdfs}.

The slope of the proton LDF is smaller than the corresponding to iron (the absolute value is greater). Then, we modify 
the slope of both LDFs such that, $\beta_{pr}(\xi)=\beta_{pr}^0-(\xi-1) \Delta \beta_0/2$ and 
$\beta_{fe}(\xi)=\beta_{fe}^0+(\xi-1) \Delta \beta_0/2$, where $\beta_{pr}^0$ and $\beta_{fe}^0$ are the proton and 
Iron slopes, respectively, obtained from the fits of Fig. \ref{FitLdfs}, $\Delta\beta_0 = \beta_{fe}^0- \beta_{pr}^0$ 
and $\xi$ is such that $\Delta \beta(\xi)=\xi \Delta \beta_0$, i.e., $\xi=1$ corresponds to the non modified case. 
Note that for $\xi=0$, $\beta_{pr}=\beta_{fe}=(\beta_{pr}+\beta_{fe})/2$. The mean and the variance of $S_b$ are 
calculated using the same procedure as before, but the previous $S(E[r_i])$ values are now modified by the factor 
$N(r_M, r_0, r_s, \beta_0) / N(r_M, r_0, r_s, \beta)$.

\begin{figure}[!bt]
\begin{center}
\includegraphics[width=10cm]{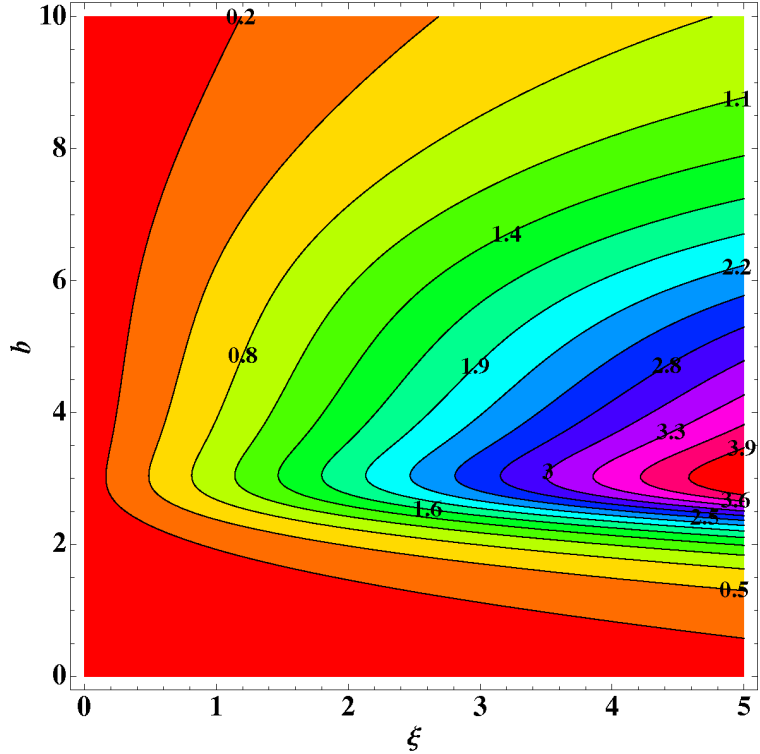}
\caption{Contour plot of $\eta(\xi,b)/\eta(1,3)$.
\label{EtaVsDbetaB}}
\end{center}
\end{figure}

Fig. \ref{EtaVsDbetaB} shows a contour plot of $\eta(\xi,b)/\eta(1,3)$ from where it can be seen that as $\xi$ increases 
$\eta$ also increases. We also see that the maximum of $\eta$ remains close to $b=3$ almost independent of $\xi$. 

\subsection{Modifying the muon content of the simulated showers}\label{ChangingMuons}

There is experimental evidence of a deficit in the muon content of the simulated showers \cite{AugerEngel-HM, 
AugerCompositionSDICRC09}. It is believed that such deficit is originated in the high energy hadronic interaction 
models which are extrapolations, over several orders of magnitude, of lower energy accelerator data. As mentioned, 
the total signal can be decomposed in the muon and electromagnetic signal. Therefore, to study how $S_b$ changes as 
a function of the muon content of the showers we modify the total LDFs in the following way: $S(r)=S_{em}(r)+f S_\mu(r)$ 
where $f$ parametrizes the artificial variation in the muon component. 

The mean and the variance of $S_b$ are calculated following the same approximations explained before. For example, the mean 
value is given by,
\begin{eqnarray}
\label{SignalEMmu} 
E[S_b] &=& \sum_{i=1}^N E \left[ S(r_i) \left( \frac{r_i}{r_0} \right)^b \right]  \nonumber \\%
&=& \sum_{i=1}^N E \left[ \left( S_{em}(r_i) +f S_{\mu}(r_i) \right) \left( \frac{r_i}{r_0} \right)^b \right]  \nonumber \\%
&\cong& \sum_{i=1}^N ( S_{em}(E[r_i]) +f S_{\mu}(E[r_i]) ) \left( \frac{E[r_i]}{r_0} \right)^b,
\end{eqnarray}
and similarly for the variance. The signal of the electromagnetic and the muonic components were also fitted separately, 
see Fig. \ref{FitLdfs}.

\begin{figure}[!bt]
\begin{center}
\includegraphics[width=10cm]{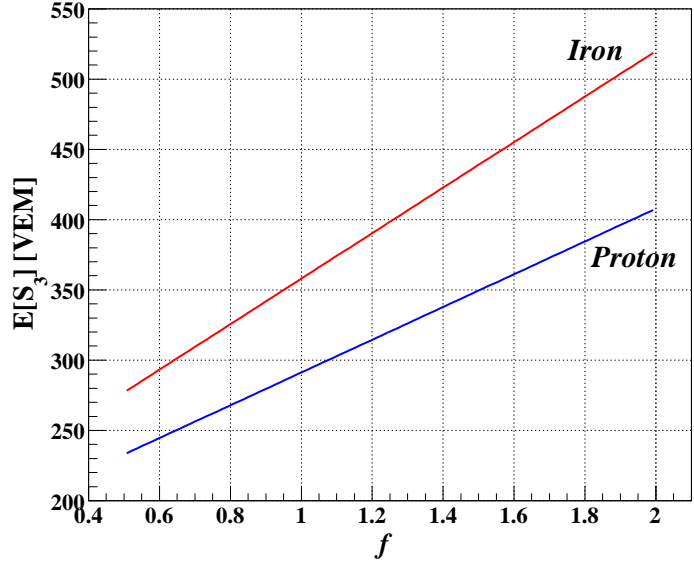}
\caption{Mean values of $S_3$ for protons and iron nuclei as a function of $f$ where $f=1$ corresponds to
the muon content predicted by QGSJET-II.
\label{S3VsFmu}}
\end{center}
\end{figure}

Fig. \ref{S3VsFmu} shows the mean values of $S_3$ for protons and Iron nuclei as a function of $f$. As expected, they 
increase with $f$. We also see that the iron curve increases faster than the proton one, which means that, for larger values 
of $f$, the discrimination power of $S_3$ also increases. This happens because the muon content of the showers is very 
sensitive to the primary mass. Then, for larger values of $f$ the muon component becomes more important increasing the mass 
sensitivity of $S_3$.

\begin{figure}[!bt]
\begin{center}
\includegraphics[width=10cm]{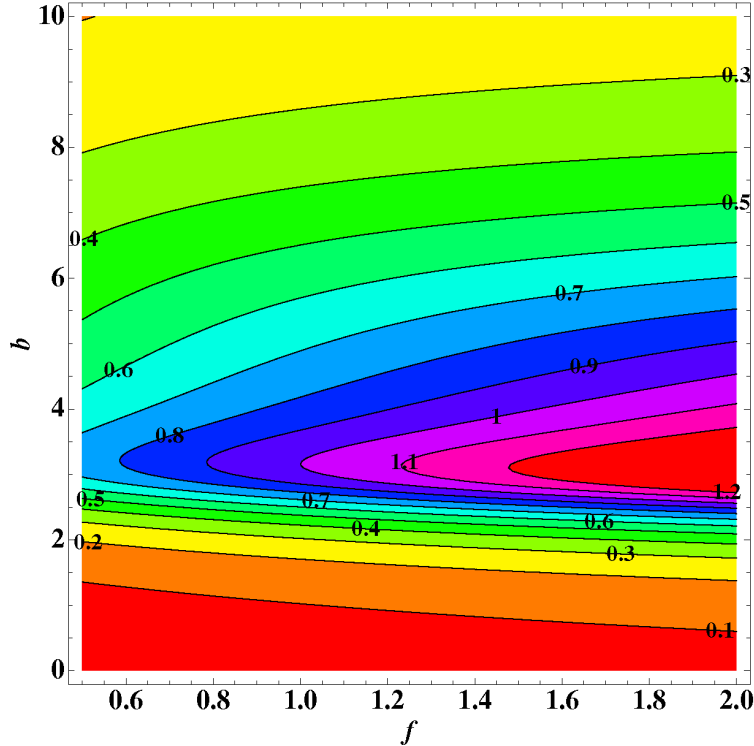}
\caption{Contour plot of $\eta(f,b)/\eta(1,3)$. $f=1$ corresponds to the muon content of the showers predicted by 
QGSJET-II.
\label{EtaVsFmuB}}
\end{center}
\end{figure}

Fig. \ref{EtaVsFmuB} shows a contour plot of $\eta(f,b)/\eta(1,3)$ from which we see how the discrimination power 
of $S_b$ increases with the muon content of the showers and that the maximum is reached at $b\cong3$ almost independently 
of $f$.

\section{Numerical analysis} \label{Numerical_Analysis}

\subsection{$Sb$ under different array geometries} \label{array_geometries}

The detection of the extensive air showers by a surface array of water Cherenkov tanks has been simulated by using our own simulation program previously reported in \cite{GRos_ropt}. This program allows us to change easily the geometry of the array and the distance between detectors. Thus, triangular and square grids have been considered varying the array spacing ($\Delta$) from 500 to 2000 meters. The injected LDF used to assign the signal to each detector is given by Eq. (\ref{NKGldf}), where $S_0$ and $\beta$ for both primaries are taken from the fits shown in Fig. \ref{FitLdfs}. Thus, we assume here almost vertical showers with energy around $10^{19.0}$ eV.

\begin{figure}[!bt]
\centerline{
\subfigure{\includegraphics[width=7.5cm]{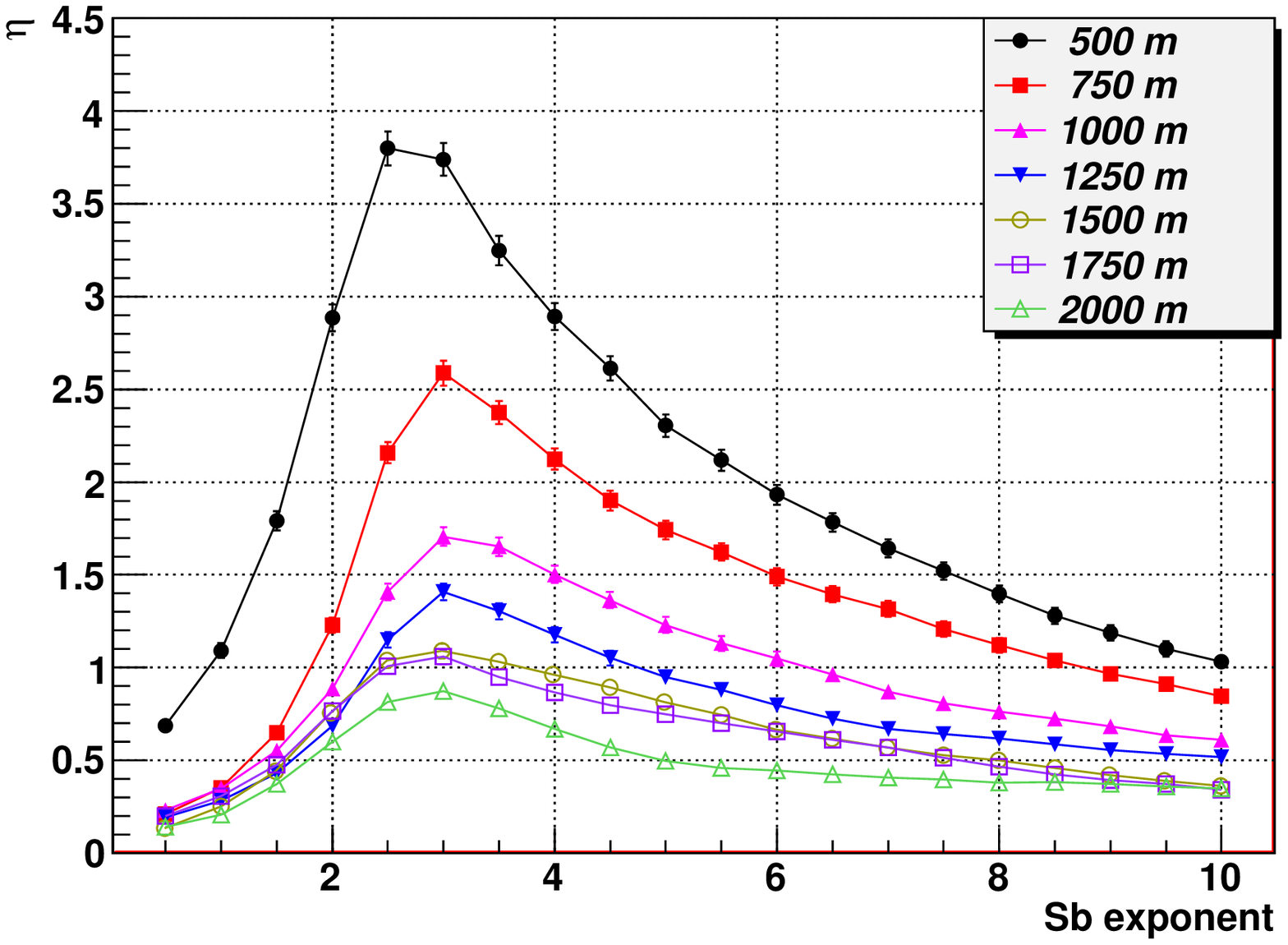} \label{fig:Merit_vs_Spacing_TrigArray}}
\hfil
\subfigure{\includegraphics[width=7.5cm]{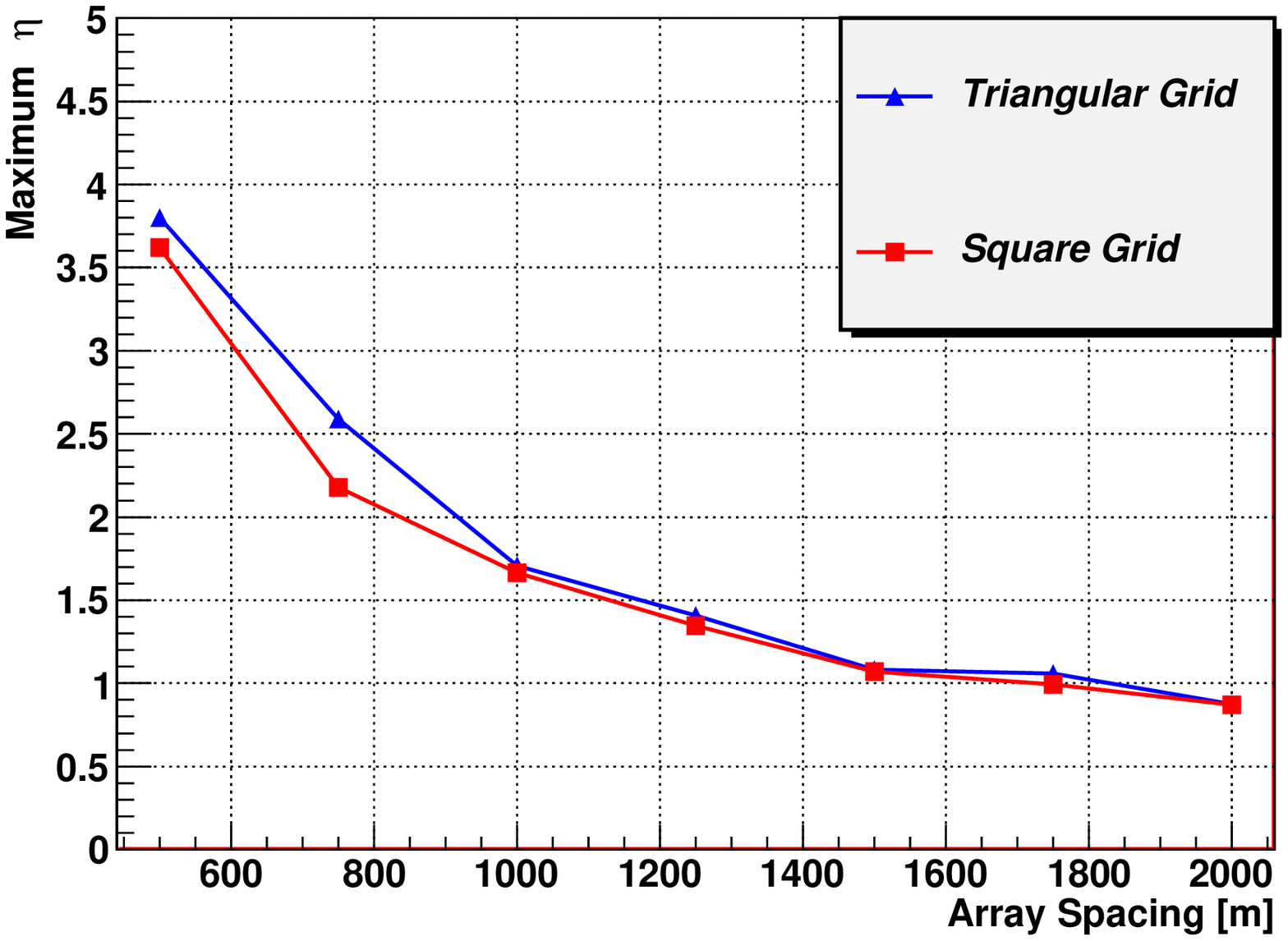} \label{fig:MaximumMerit_vs_spacing}}
}
\caption{Left: Merit factor as a function of $S_b$ exponent for several distance between detectors in a triangular grid. Right: $S_b$ merit
factor for the optimum $b$ as a function of the array spacing for a triangular and a square grid.}
\label{fig:b_vs_spacing}
\end{figure}

We use again the merit factor, $\eta$ (see Eq. (\ref{Sb})), to quantify the discrimination power of $S_b$ as a 
function of the exponent $b$ but, contrary to the previous Section, we prefer now to use the median and $\sigma_{68}^2$ instead of the mean value and the variance, respectively. The result for a triangular grid is shown in Fig. \ref {fig:Merit_vs_Spacing_TrigArray}. It can be seen that, despite the fact that the value of $\eta$ decreases markedly as the density of the array diminishes, the maximum discrimination power is reached for $b=3$ almost independently of the array spacing. The result is in agreement with the analytical approach for the 1500 m array (see Fig. \ref{EtaVsB}), although the merit factor is now smaller because shower fluctuations are also taken into account in the present calculation. Similar results have been obtained in the case of a square grid, showing that  $S_b$ encodes mainly information about the shower front structure and not just about the array geometry.

The maximum merit factor as a function of the array spacing is shown in Fig. \ref{fig:MaximumMerit_vs_spacing}
for both the triangular and square array. The merit factor increases as the array spacing decreases, as it was expected since the LDF is sampled in more points as the array becomes denser, improving $S_b$ separation power. In case of $\Delta < 1000$ m the merit factor is larger for the triangular grid since the number of triggered stations is also larger for this geometry.


\subsection{Simulations} \label{Simulations}

In what follows, we perform a more realistic simulations in order to treat more accurately the tank response and 
to take into account experimental uncertainties such as the shower reconstruction and the hadronic interaction model. In addition, we extend the energy and zenith angle range.

The simulation of atmospheric showers is performed by using the AIRES Monte Carlo program (version 2.8.4a) 
\cite{Sciutto} with QGSJET-II \cite{QGSJet} and Sibyll 2.1 \cite{Sibyll} as the hadronic interaction models (HIM). Since  the number of secondary particles produced in a shower is extremely large (i.e. $\sim10^{11}$ particles in a proton shower of $10^{20}$ eV), it is very costly, in processing time and disk space, to follow all of them. Therefore, we use a statistical method called thinning, first introduced by M. Hillas \cite{Thinning1}, as it is implemented in AIRES. A relative thinning of $10^{-6}$ and weight factor of 0.2 are used for the generation of the showers. Iron and proton primaries are simulated in the energy range from $10^{19}$ to $10^{19.6}$ eV and the arrival direction follows an isotropic distribution with zenith angle in the range $0^\circ \leq \theta \leq 60^\circ$. The number of simulated showers per HIM and primary corresponds to an exposure of 5165 $km^2 sr yr$ ($\sim$ 0.8 years of Auger full operation \cite{AugerSDSpectrumICRC07}).

The simulation of the response of the surface detectors as well as the shower reconstruction are performed using the official Offline reconstruction framework of the Pierre Auger Observatory \cite{Offline}. The simulation includes a triangular grid of Cherenkov detectors of 1.5 km of spacing. The unthinning method of P. Billoir \cite{Billoir:08} is used to obtain a list of particles that hit a given detector of the array. The GEANT4 package \cite{Geant4} is used to simulate the behavior of particles inside the tanks. The surface detector simulation has been tested and proved to be in good agreement with experimental data \cite{GhiaSDSims}. In order to increase the statistics, each shower is recycled 5 times by randomly distributing their cores inside the array (full discussion of the statistical effects of recycling air showers could be found in \cite{Repetitions:08}).

The reconstructed energy has been simulated by fluctuating the real one using a Gaussian uncertainty of $20\%$, a typical value for surface 
experiments \cite{AGASASpectrum, AugerSDSpectrumICRC07}.

Simulations are divided in logarithmic energy bins from $\log(E/eV)=19$ to $\log(E/eV)=19.6$ in steps of $0.1$. 
We also consider three different bins of zenith angles centered at 30$^\circ$, 45$^\circ$, and 55$^\circ$ and of 
$10^\circ$ wide.

\subsection{Optimization}

\begin{figure}[!bt]
\begin{center}
\includegraphics[width=10cm]{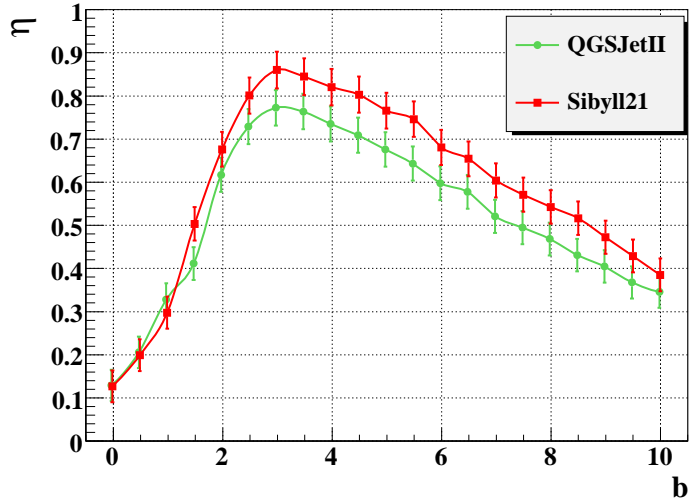}
\caption {Merit factor of $S_b$ as a function of $b$ obtained from simulated events. There is a general agreement with 
Fig. \ref{EtaVsB} and Fig. \ref{fig:Merit_vs_Spacing_TrigArray}}
\label {fig:Sim_eta_vs_exp}
\end{center}
\end{figure}

Fig. \ref{fig:Sim_eta_vs_exp} shows the merit factor of $S_b$ as a function of $b$ for both HIM considered. The maximum is reached at $b\cong3$ in a very good agreement with the result obtained in the analytical analysis (see Fig. \ref{EtaVsB}). Furthermore, the shape of the curve is quite similar. Nevertheless, the merit factor is lower and the peak wider, as it is expected because every event, independently of its energy or zenith angle, has been included and shower fluctuations and the effects introduced by the detectors and reconstruction methods are accounted for. The result is also in agreement with previous numerical simulation (Fig. \ref{fig:Merit_vs_Spacing_TrigArray}). 

\subsection{Influence of distant detectors}

Several tests have been performed to study whether distant stations from shower axis, for which the fluctuations in the 
total signal are quite significant, could affect the separation power of $S_3$. Note that in the $S_3$ sum, saturated 
stations are not included. 

\begin{figure}[!bt]
\begin{center}
\includegraphics[width=10cm]{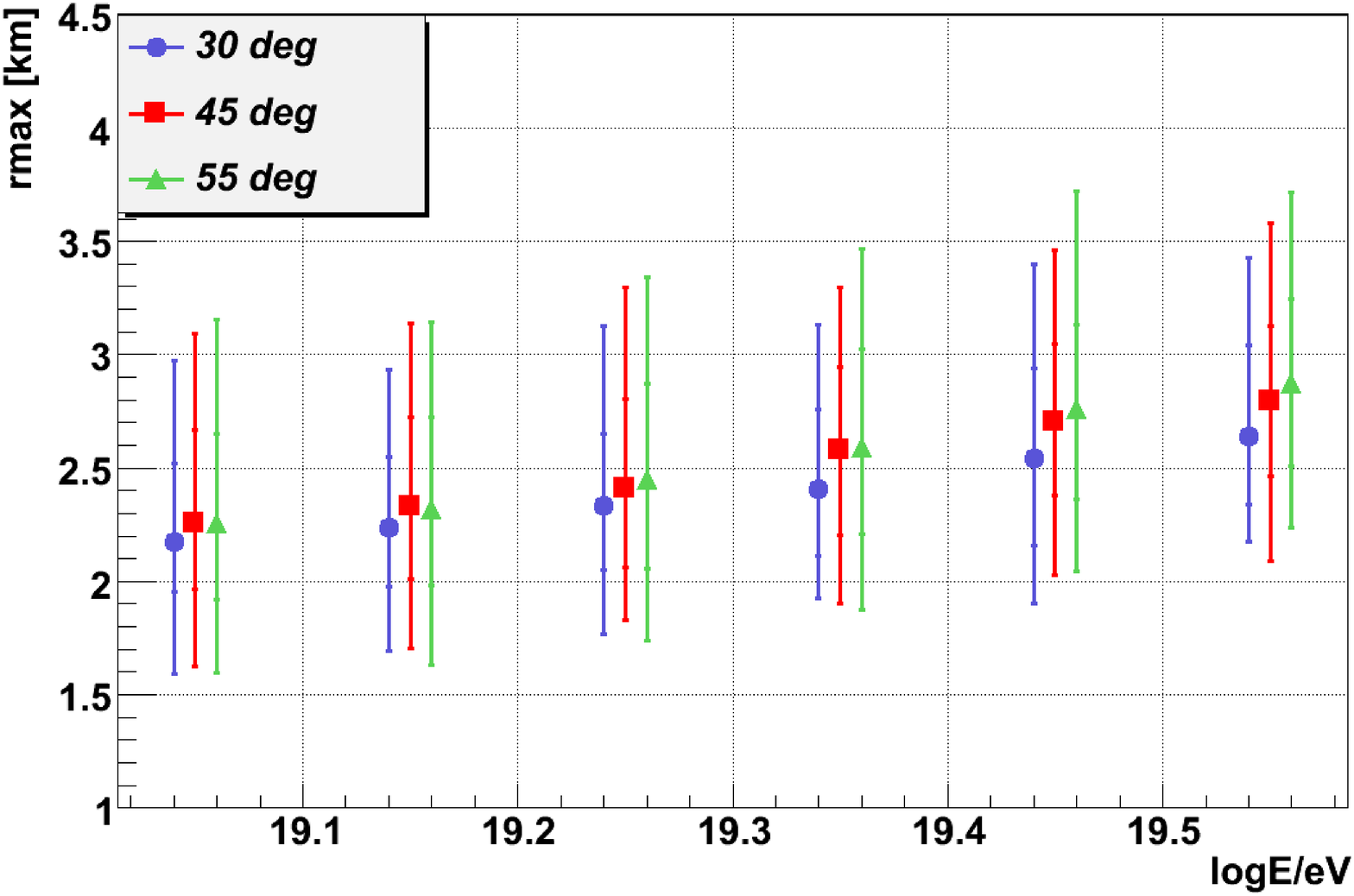}
\caption{Distance from the furthest triggered station to shower axis for three different zenith angles
as a function of energy. Both hadronic interaction models and primaries are included. The error bars correspond to the
68\% and 95\% confidence levels.}
\label{fig:rmax}
\end{center}
\end{figure}

Let us call $r_{lim}$ to the maximum distance of the stations considered in the $S_3$ sum. In previous analysis, no cut in 
distance were applied so that all the triggered detectors were included with the same weight. Now, different cuts are tested:
\begin{itemize}

\item $r_{lim}=r_{opt}$. We call $r_{opt}$ to the distance obtained by searching the value that maximizes the merit factor 
in 100 m steps. It depends on the hadronic interaction model and zenith angle.

\item $r_{lim} = 2700 m$. All the stations at a distance from shower core larger than 2700 m are excluded. This value 
is selected, somehow arbitrarily, in order to reject the stations whose signal could be dominated by fluctuations.

\item $r_{lim} = r_{max}(E,\theta)$. Considering the distribution of the distance from the furthest triggered detector to 
the shower axis (see Fig. \ref{fig:rmax}), $r_{max}$ is defined as the median of such distribution. In the calculation for each 
energy and zenith angle intervals, proton and Iron primaries and both HIM are included.

\item Each term of the $S_3$ sum is weighted using the so called Lateral Trigger Probability (LTP), i.e. the probability 
of a shower to trigger a detector as a function of the reconstructed energy, shower zenith angle and the distance from the 
detector to the shower axis. This is an elegant way of switching off smoothly the stations at large distances from the core. 
The LTP is obtained as described in \cite{LTP-MCMedina}. 

\item $r_{lim} \rightarrow \infty$. All the triggered stations are included. No cut is applied.

\end{itemize}

\begin{figure}[!bt]
\begin{center}
\includegraphics[width=10cm]{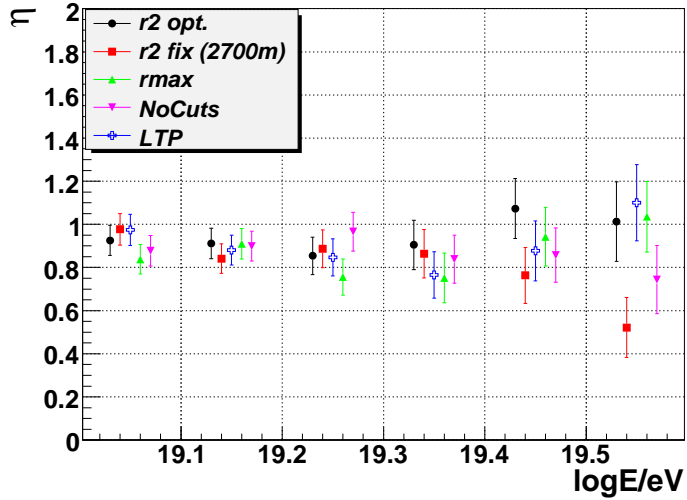}
\end{center}
\caption{Merit factor as function of the reconstructed energy. Several cuts are tested (see the text for details). Here the result 
for Sibyll 2.1 and $\theta = 55\pm5^\circ$ is shown. Similar result are obtained for QGSJET-II and other zenith angle bins.}
\label{fig:Merit_Cuts}
\end{figure}

Fig. \ref{fig:Merit_Cuts} shows $\eta$ as a function of the primary energy for the different cuts mentioned before. It can be 
seen that the difference among them is not significant, showing the robustness of the parameter. Therefore, the discrimination 
power of $S_3$ is not affected by distant stations where the fluctuations could dominate the signal. Since the cuts do not improve 
the parameter ability, we recommend no to use a cut in distance. Furthermore, these cuts could introduce bias in composition 
determination since the LDF of a proton primary is steeper than that of an Iron nuclei of the same energy and zenith angle 
(see Fig \ref{FitLdfs}).

\subsection{$S_3$ dependence on the primary energy and zenith angle}

\begin{figure}[!bt]
\begin{center}
\includegraphics[width=10cm]{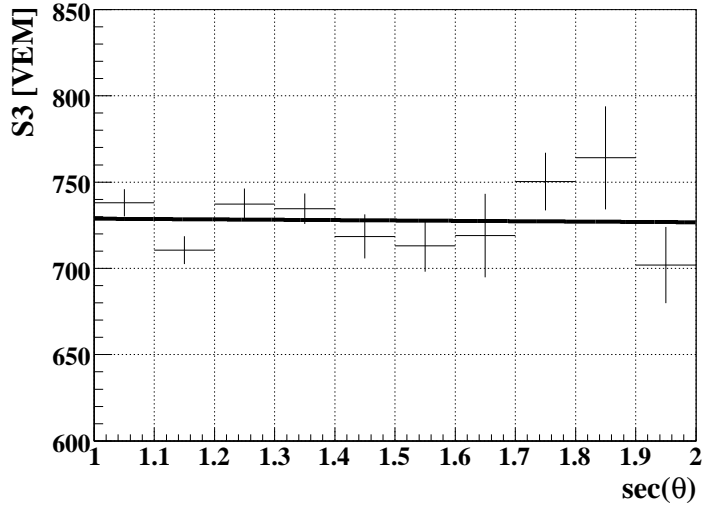}
\end{center}
\caption{$S_3$ vs. $sec(\theta)$ for proton primaries (Sibyll 2.1). A linear fit to the points shows a negligible dependence 
on the zenith angle of the incoming shower. Error bars are $RMS/\sqrt{N}$.}
\label{fig:S3_vs_SecTheta}
\end{figure}

Fig. \ref{fig:S3_vs_SecTheta} shows the mean value of $S_3$ as a function of $\sec(\theta)$ for protons primaries and Sibyll 2.1. 
There is no significant dependence with the zenith angle. Similarly occurs in case of Iron primaries and QGSJET-II.

\begin{figure}[!bt]
\centerline{
\subfigure{\includegraphics[width=7cm]{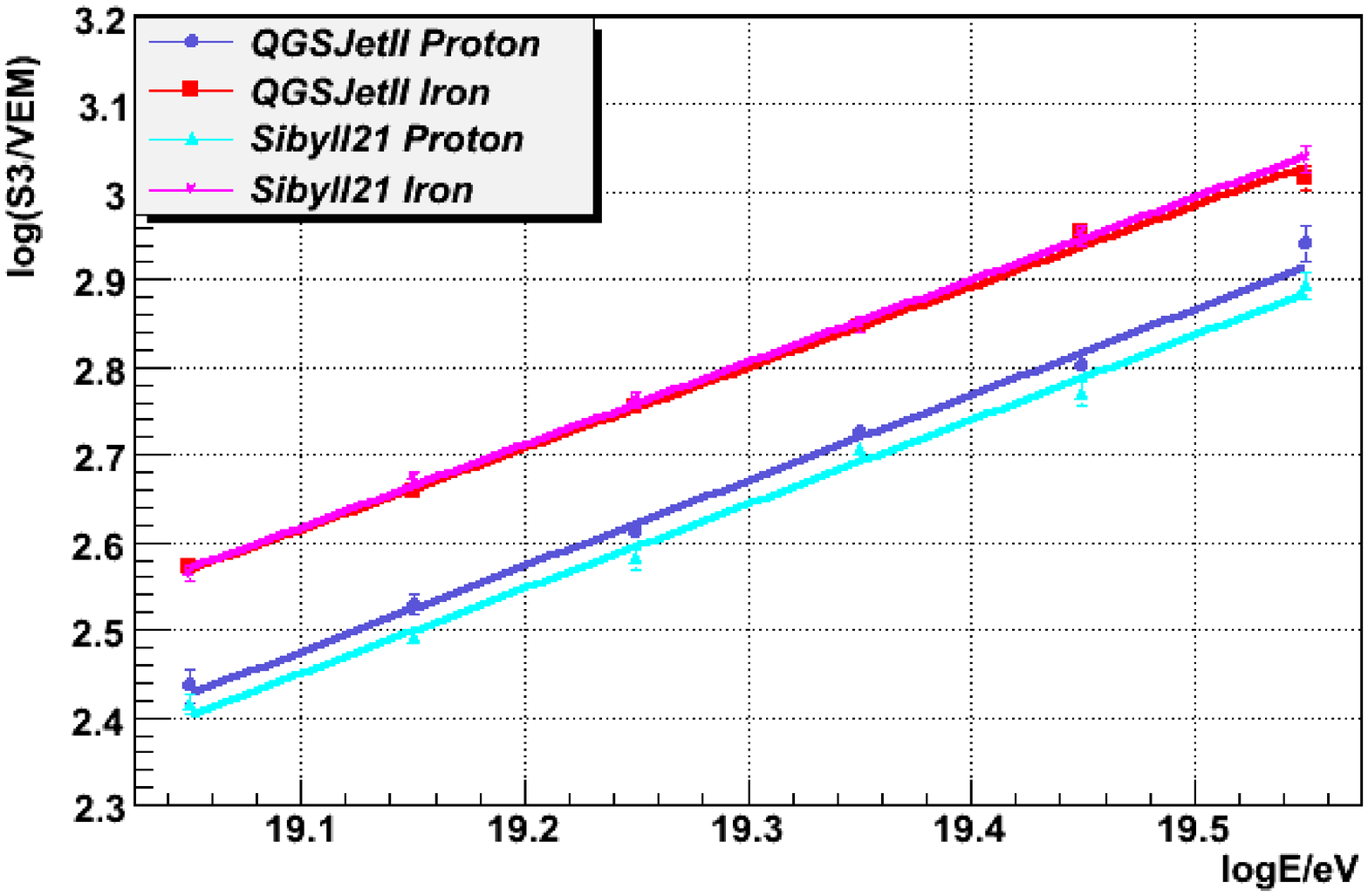} \label{fig:logS3_vs_logE_30deg}}
\hfil
\subfigure{\includegraphics[width=7cm]{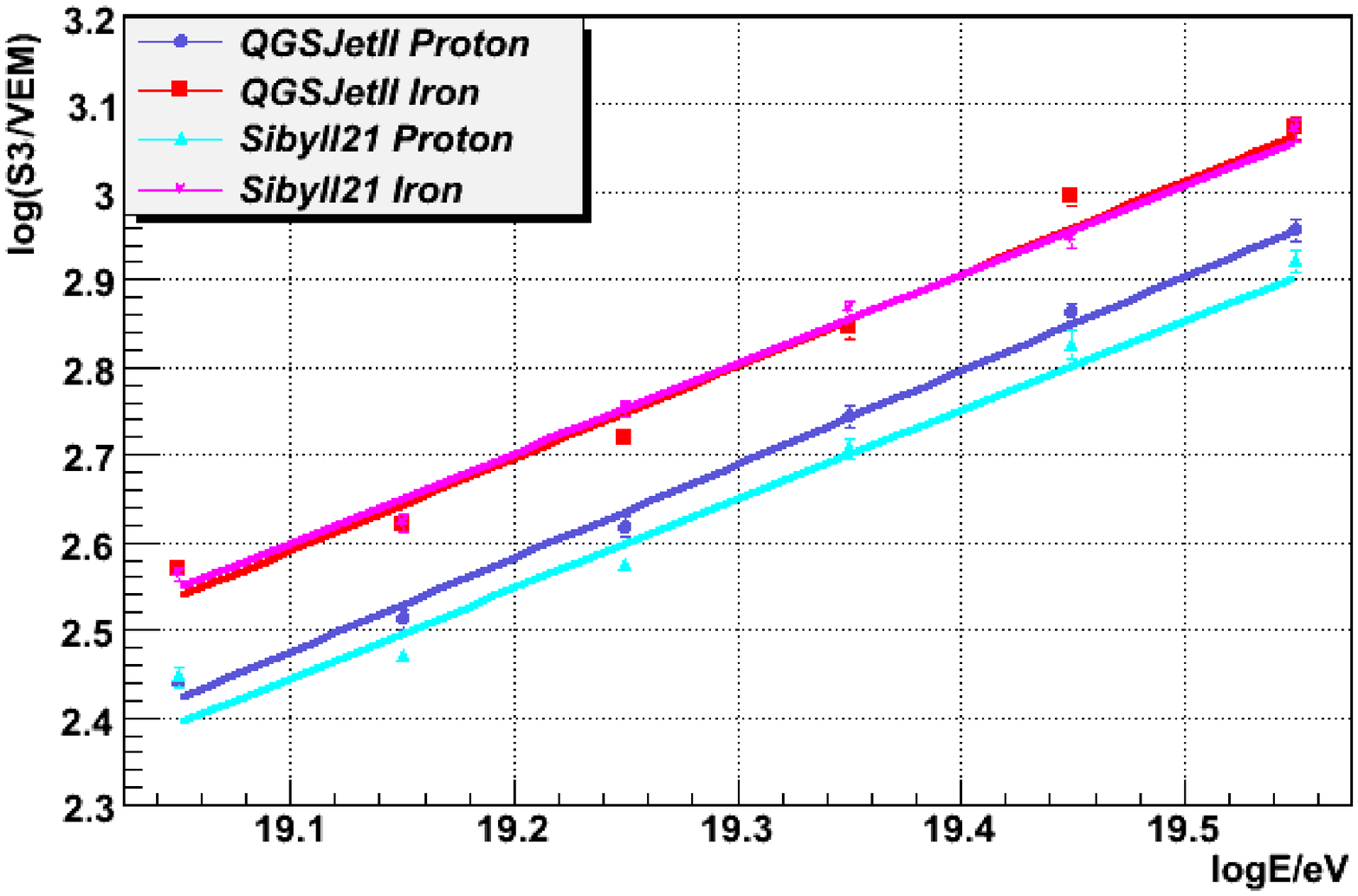} \label{fig:logS3_vs_logE_45deg}}
}
\caption{$\log(S_3/VEM)$ vs. $\log(E/eV)$ for $\theta=30^\circ$ (left) and $\theta=45^\circ$ (right). The error bars 
are the $RMS/\sqrt{N}$. Linear fits to the points are also shown.}
\label{fig:S3_vs_logE}
\end{figure}

The evolution of $S_3$ with energy for proton and iron primaries is shown in Fig. \ref{fig:S3_vs_logE} for two zenith
angle bins, centered at $\theta = 30^\circ$ and $\theta = 45^\circ$ and $10^\circ$ wide. An almost linear dependence has been found.

\section{A realistic application} \label{Application}

The reliability of the composition determination by using $S_3$ and other mass sensitive parameters is checked in this Section. 
We select the two parameters most commonly used in the literature, one from the surface detectors, the rise time at 1000 m 
from shower core, and the other from fluorescence telescopes, $X_{max}$, the atmospheric depth at which the maximum development 
of the cascade is reached. A brief discussion follows on specific details about the determination of these two parameters:

\begin{itemize}

\item Rise time at $r_0=1000$ m from core, $t_{1/2}(r_0)$ [ns]: the rise time of a single station is defined as the time it takes 
to increase from $10\%$ to $50\%$ of the total signal. The spread of the arrival times of the shower particles at a 
fixed core distance increases for smaller production heights, so the rise time is expected to be smaller for heavy primaries that develop higher in the atmosphere. 
$t_{1/2}(r_0)$ is obtained by fitting the rise time of each triggered station using the function $t_{1/2}(r)=(40+ar+br^2)$ ns and evaluating the fitting function at
$r_0=1000$ m. Parameters \emph{a} and \emph{b} are free in the fit. Only stations whose distance to shower axis is in the range from 600 to 1500 m and whose signal is 
greater than 10 VEM are included in this fit in order to avoid signals dominated by large fluctuations and saturated ones. Since three stations are required at least in the fit, 
the number of events for which is possible to evaluate $t_{1/2}(r_0)$ is significantly reduced, specially at larger zenith angles.

\item $X_{max}$ [$g/cm^2$]: up to now, $X_{max}$ is regarded as the most useful parameter for composition analysis. $X_{max}$
is very sensitive to the identity of the primary and less affected by uncertainties in energy determination compared to
surface parameters, since it depends logarithmically on primary energy. In order to calculate a realistic $X_{max}$, 
we use the value calculated by AIRES and fluctuate it by using a Gaussian distribution whose standard deviation is 
a typical value for the resolution achieved in fluorescence experiments \cite{AugerXmax2010,HiResElongationRate},
accounting for the response of the detector and the effects of the reconstruction method.

\end{itemize}

The parameters $S_3$ and $X_{max}$ are almost independent of zenith angle, so it is possible to combine events
of different zenith angles in a given sample. Obviously, this is not the situation for $t_{1/2}(r_0)$. To compensate 
this dependence a quadratic fit is performed $t_{1/2}(r_0)$ vs. $sec(\theta)$ for each primary and hadronic model, and 
using average values of the fitted parameters, we correct, in a simple way, for the zenith angle dependence of $t_{1/2}(r_0)$:
\begin{equation}\label{eq:risetime_correction}
t_{1/2}^{corr}(r_0,\sec\theta) = t_{1/2}^{meas}(r_0,\sec\theta) +%
\left(t_{1/2}^{fit}(r_0,1.05)-t_{1/2}^{fit}(r_0,\sec\theta)\right).
\end{equation}
The correction does not increase the fluctuations and $t_{1/2}^{corr}(r_0)$ shows a strong reduction on the zenith angle 
dependence. 

For the subsequent analysis, we consider the lowest energy bin (form $10^{19}$ to $10^{19.1}$ eV) where we have the largest number 
of events. The samples corresponding to a given primary, HIM and one of the mass sensitive parameters considered are binned. Let 
us call $h_{p}(i)$ and $h_{fe}(i)$ to the number of events in the $i$th bin, normalized to the total number of events in the sample, 
corresponding to a given parameter and for protons and iron nuclei, respectively. The histograms $h_{p}$ and $h_{fe}$ are assumed 
to be the distribution of the universe. The proton abundance or composition of a sample is defined as $C_p=N_{p}/(N_{p}+N_{fe})$, 
where $N_{p}$ and $N_{fe}$ are the number of protons and iron nuclei in a given sample. We consider samples of $C_p^{true}$ from 
$0$ to $1$ in steps of $0.1$. For each value of $C_p^{true}$, we generate 500 samples of $N_{s}=300$ events each by taking at 
random values from the histograms $h_{p}$ and $h_{fe}$. For each of these new samples we generate a histogram $H_{s}$, with the 
same binning used in $h_{p}$ and $h_{fe}$, which is not normalized and then satisfies $\sum_{i} H_{s}(i) = N_{s}$. Thus, assuming 
Poisson statistics, the probability of having a configuration with $H_{s}(i)$ events in the $i$th bin is given by, 
\begin{equation}\label{poisson}
P(\{H_{s}(i)\}_{i}) = \prod_{i} \exp{(-H_{t}(i))} \times \frac{H_{t}(i)^{H_{s}(i)}}{H_{s}(i)!} 
\end{equation}
where $H_{t}(i)=N_{s}\, (C_p h_{p}(i) + (1-C_p) h_{fe}(i))$ and $C_p$ is the unknown parameter. Therefore, the inferred proton abundance 
is obtained by minimizing $-\ln P(\{H_{s}(i)\}_{i})$. 

\begin{figure}[!hbt]
\centerline{
\subfigure{\includegraphics[width=7cm]{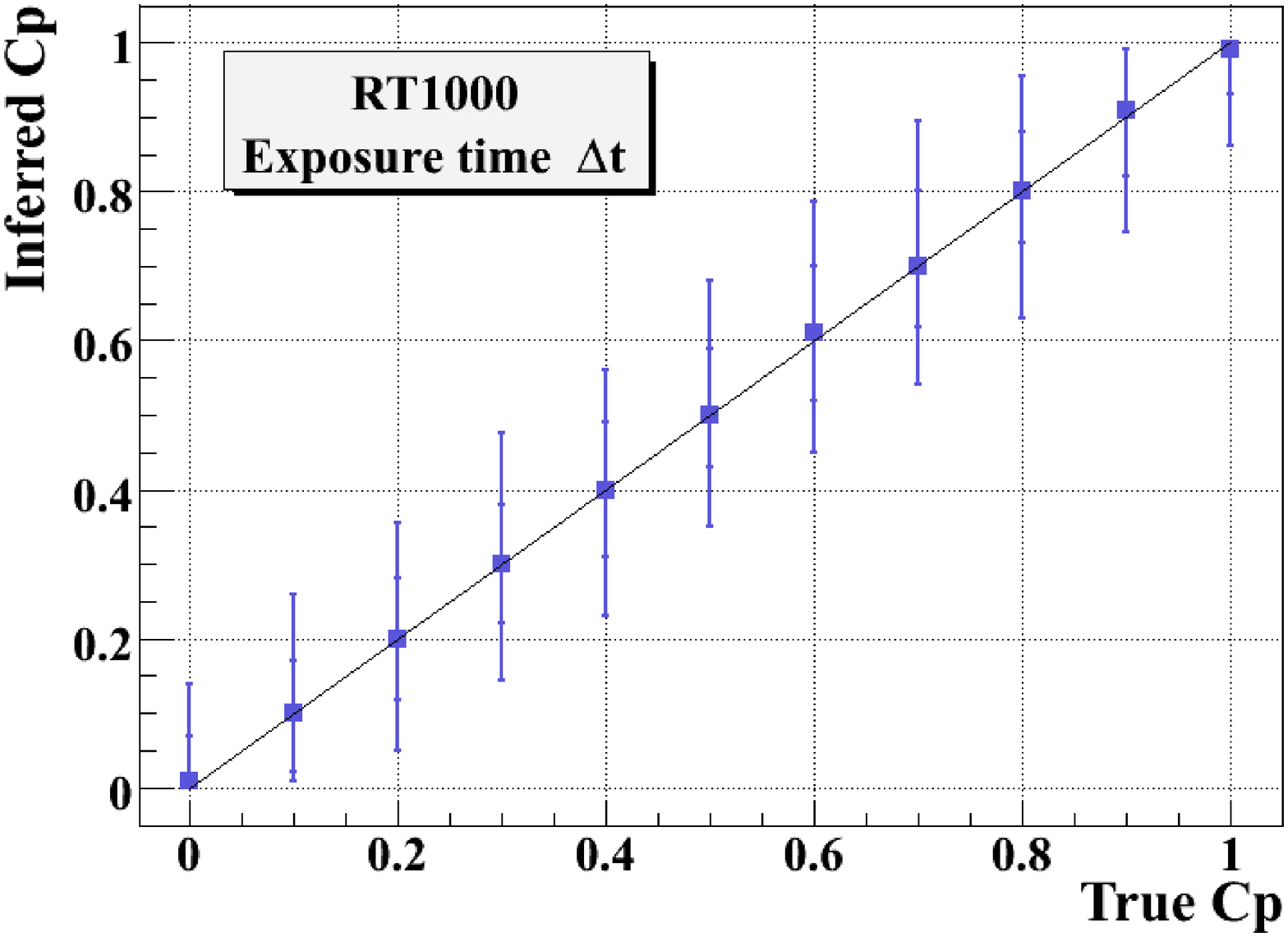} \label{fig:RT_QG}}
\hfil
\subfigure{\includegraphics[width=7cm]{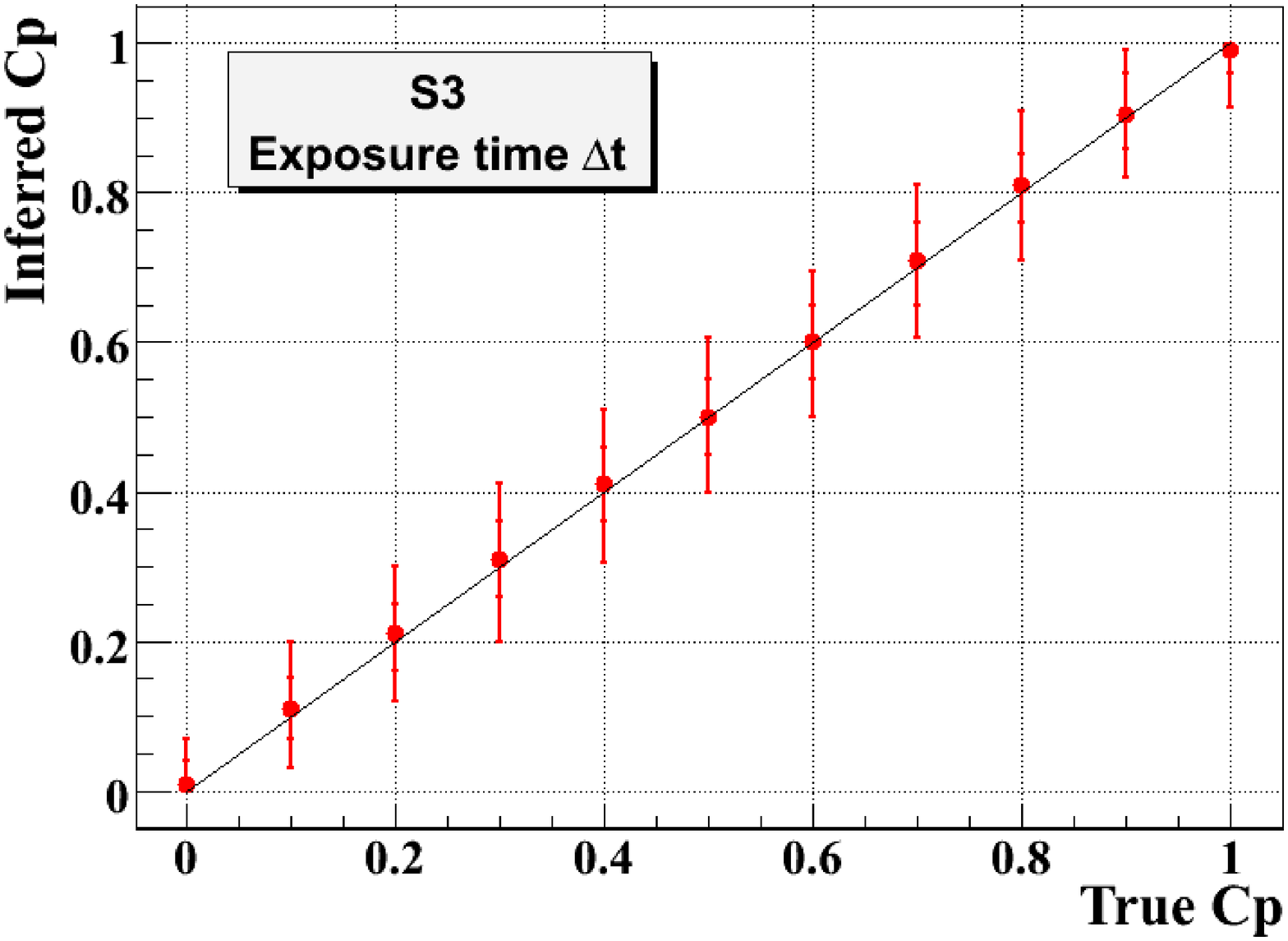} \label{fig:S3_QG}}
}
\centerline{
\subfigure{\includegraphics[width=7cm]{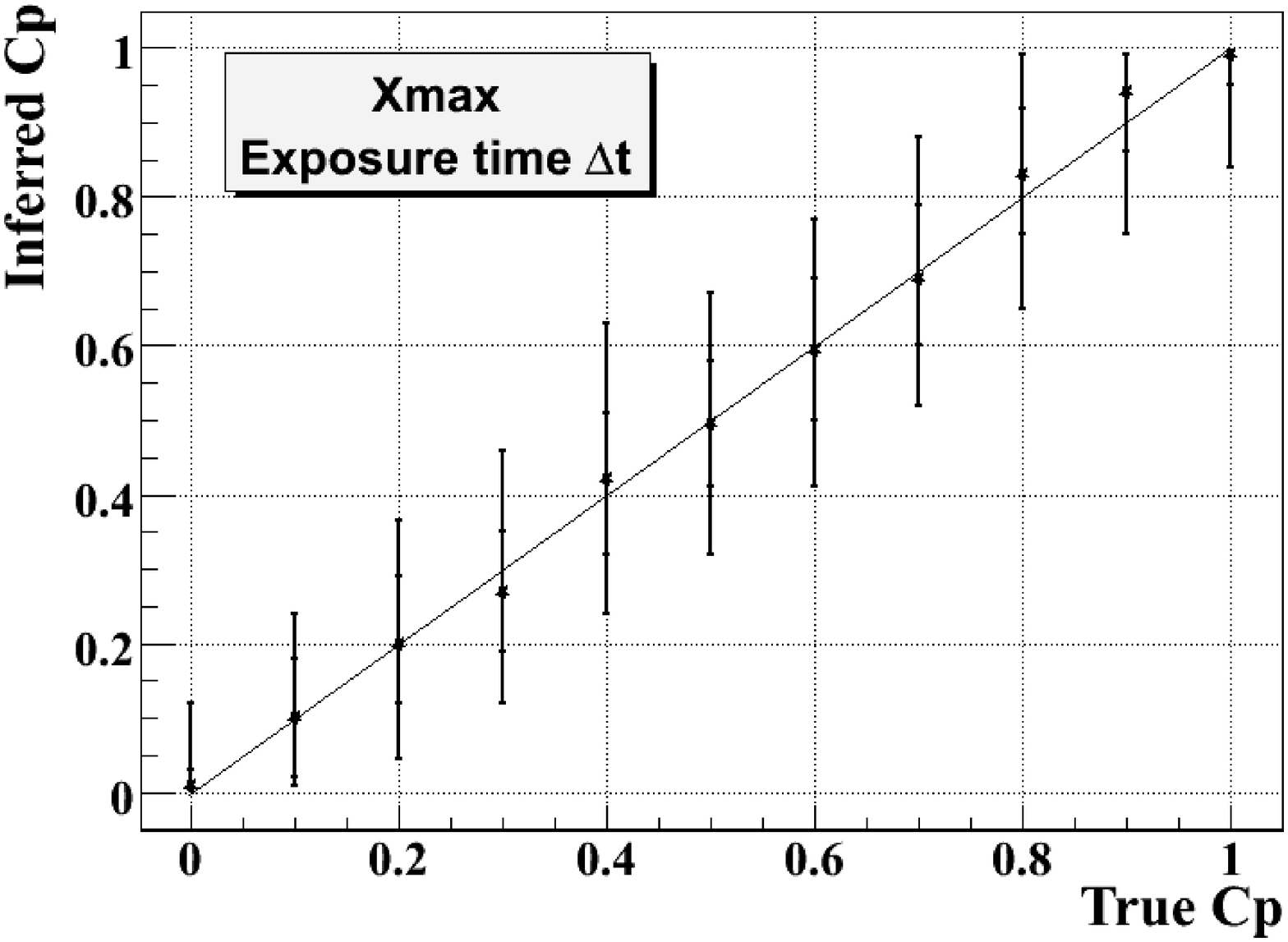} \label{fig:Xmax10pc_QG}}
\hfil
\subfigure{\includegraphics[width=7cm]{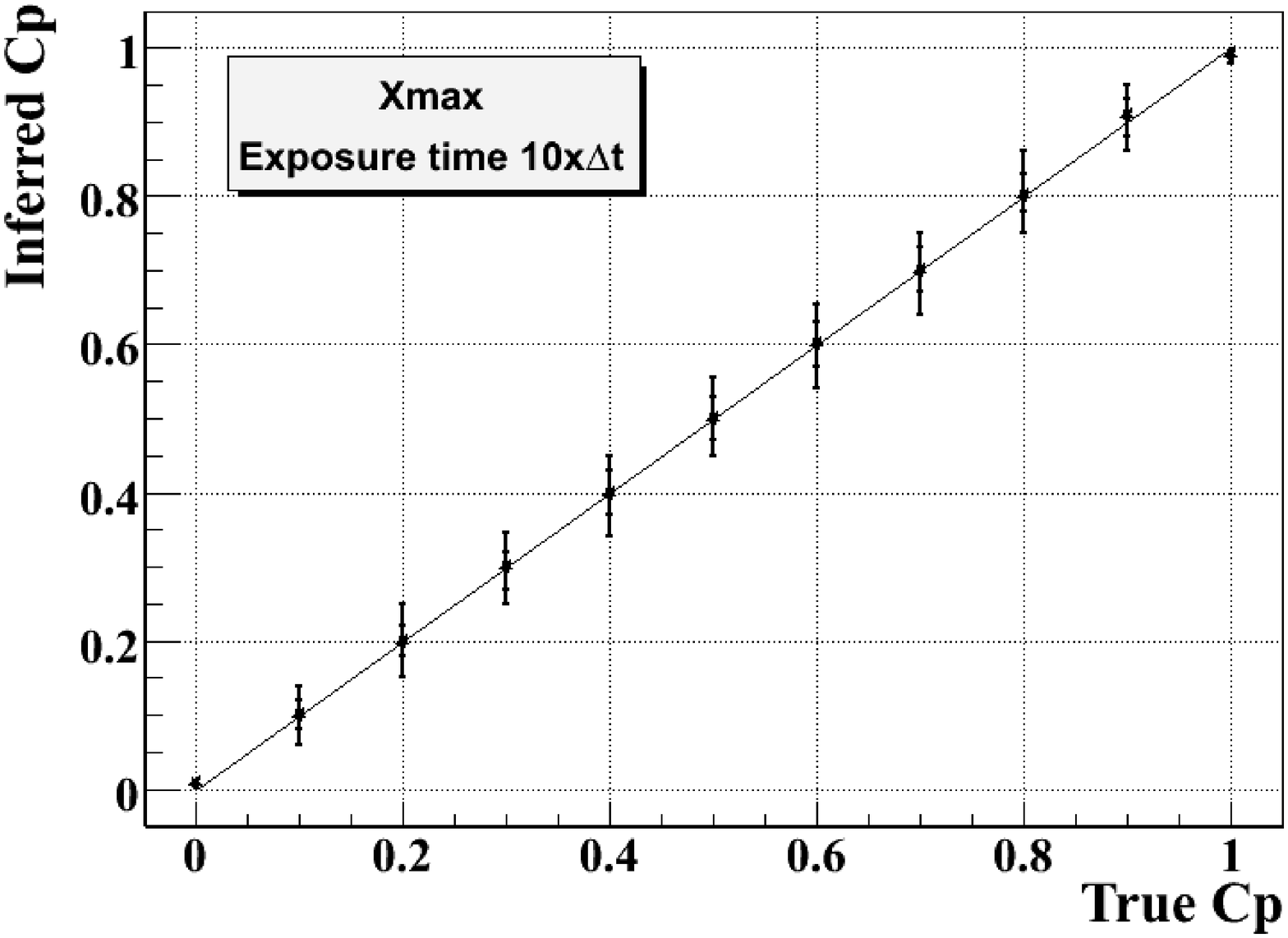} \label{fig:Xmax_QG}}
}
\caption{True vs. inferred proton fraction at $\sim 10^{19}$ eV using QGSJET-II for $t_{1/2}(r_0)$ (top-left), $S_3$ (top-right) and
$X_{max}$ (bottom-left) for a fixed exposure time (the 10\% duty cycle of the fluorescence telescopes is taking into account). The
bottom-right panel shows the inferred proton fraction obtained using $X_{max}$ but for samples 10 times larger, corresponding to the
same SD exposure.}
\label{fig:Cp}
\end{figure}

Fig. \ref{fig:Cp} shows the inferred composition as a function of the true composition corresponding to QGSJET-II. For Sibyll 2.1 
similar results are obtained but with smaller error bars (which represent the $68\%$ and $95\%$ C.L.), in agreement with the fact 
that the merit factors are in general greater for this HIM. 

The statistics available for each mass sensitive parameter corresponding to a given exposure time, is a key value to compare
their discrimination capabilities. Due to the limited duty cycle of the fluorescence telescopes, only 10\% of the events are
detected, so the statistics for $X_{max}$ is significantly lower than that for surface parameters (ground detectors have almost
full operation time). This fact has been considered in Fig. \ref{fig:Cp}. Just to illustrate the significance of taking into
account the limited statistics for $X_{max}$ when doing composition studies, it is also shown the result for $X_{max}$ if the
same statistics as the surface parameters were available. Then, the error bars are reduced becoming the smallest ones, but ten
times more exposure would be required.

A second study has been performed. Now, a fix true proton fraction $C_p^{true}=0.5$ is assumed and the inferred proton fraction 
is calculated in the energy range from $10^{19.0}$ to $10^{19.6}$ eV. In this case, in order to improve the small statistics in 
the higher energy bins, the histograms $h_{p}(i)$ and $h_{fe}(i)$ corresponding to a given parameter, energy bin and HIM are fitted. 
The fitting function used is the so-called Asymmetric Generalized Gaussian (AGG) \cite{AGG} which is defined as, 
\begin{eqnarray}\label{eq:AGG}
P_{AGG}(y) = \begin{cases}\frac{c\, \gamma_a}{\Gamma(1/c)}
\ \exp[-\gamma_l^c\ (-y+\mu)^c\, ] & if~ y < \mu\cr 
\frac{c\, \gamma_a}{\Gamma(1/c)} 
\ \exp[-\gamma_r^c\ (y-\mu)^c\, ] & if ~ y \geq \mu
\end{cases} 
\end{eqnarray}
where,
\begin{equation}
\gamma_a=\frac{1}{\sigma_l + \sigma_r} \left(
\frac{\Gamma(3/c)}{\Gamma(1/c)} \right) ^{1/2}, \; \;
\gamma_l=\frac{1}{\sigma_l} \left( \frac{\Gamma(3/c)}{\Gamma(1/c)}
\right)^{1/2}, \; \; \gamma_r=\frac{1}{\sigma_r} \left(
\frac{\Gamma(3/c)}{\Gamma(1/c)} \right)^{1/2}, \\ \nonumber
\end{equation}
$\sigma_l^2$ and $\sigma_r^2$ are the variances of the left and right sides of the probability density function, 
respectively, and $\Gamma(x)$ is the Gamma function. If $\sigma_l^2 = \sigma_r^2$, the AGG is symmetric. Furthermore, if 
$\sigma_l^2 = \sigma_r^2$ and $c=2$, AGG reduces to the regular Gaussian distribution function and, for $c=1$, it 
represents the Laplace distribution.

\begin{figure}[!bt]
\centerline{
\subfigure{\includegraphics[width=7cm]{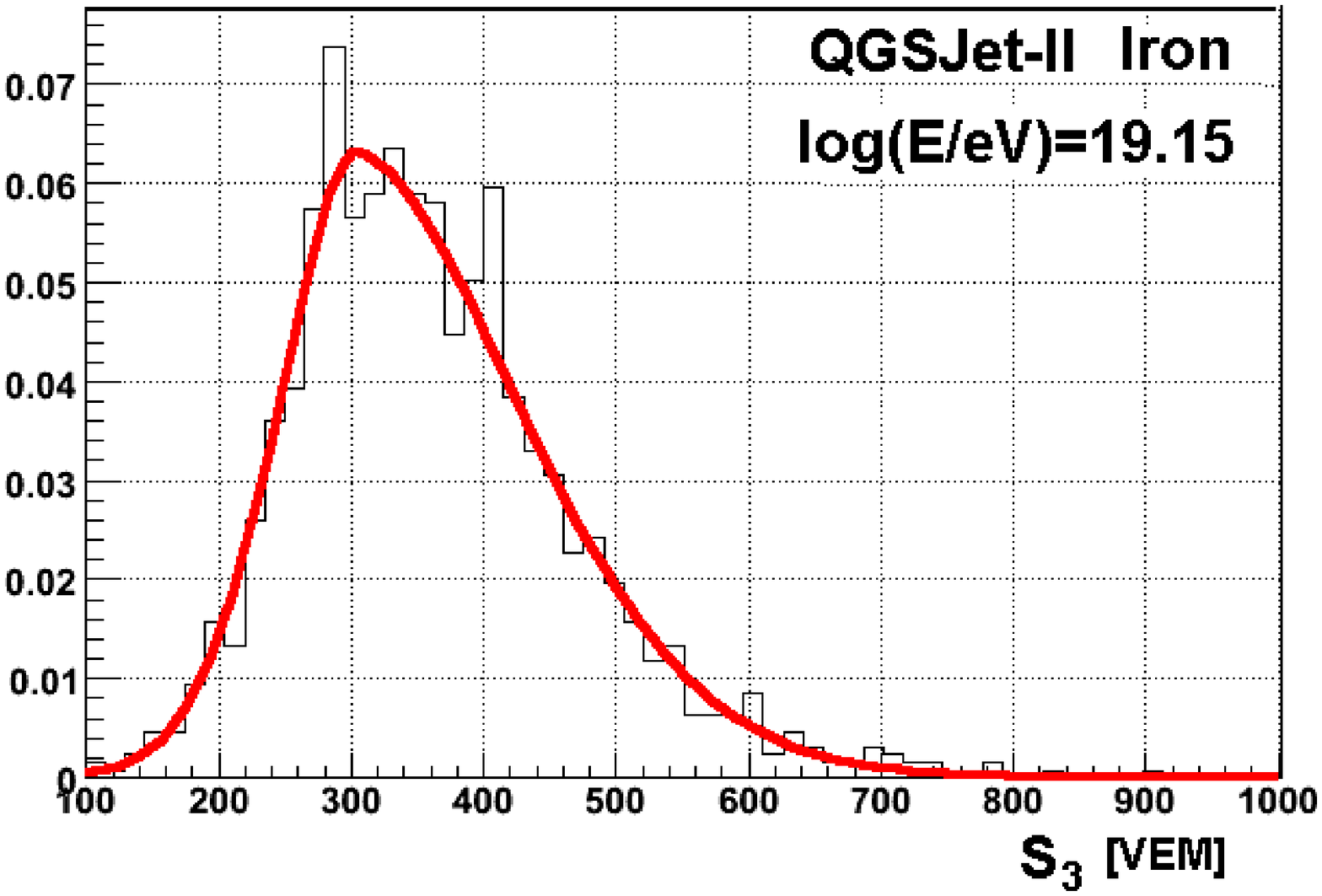}}
\hfil
\subfigure{\includegraphics[width=7cm]{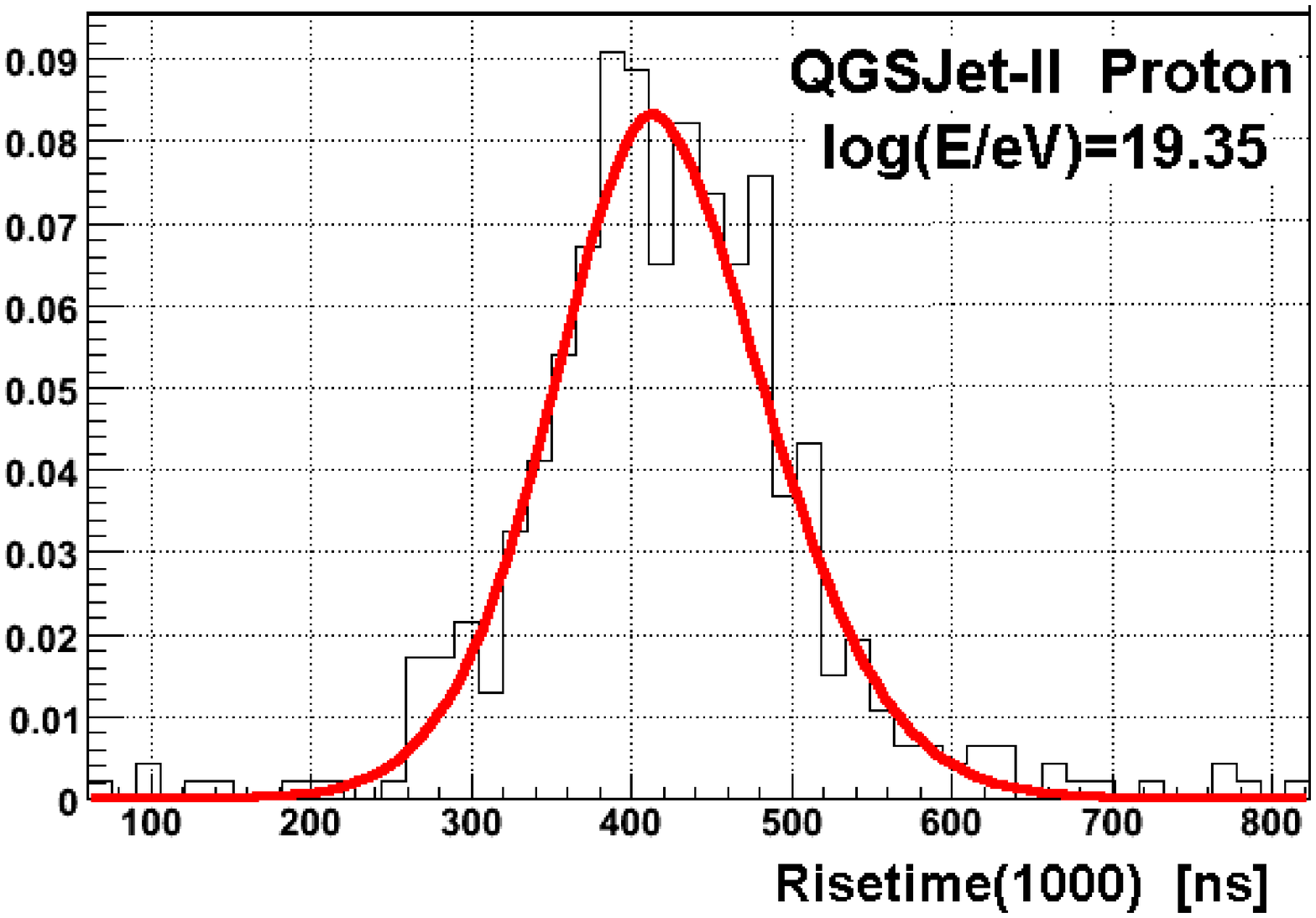}}
}
\centerline{
\subfigure{\includegraphics[width=7cm]{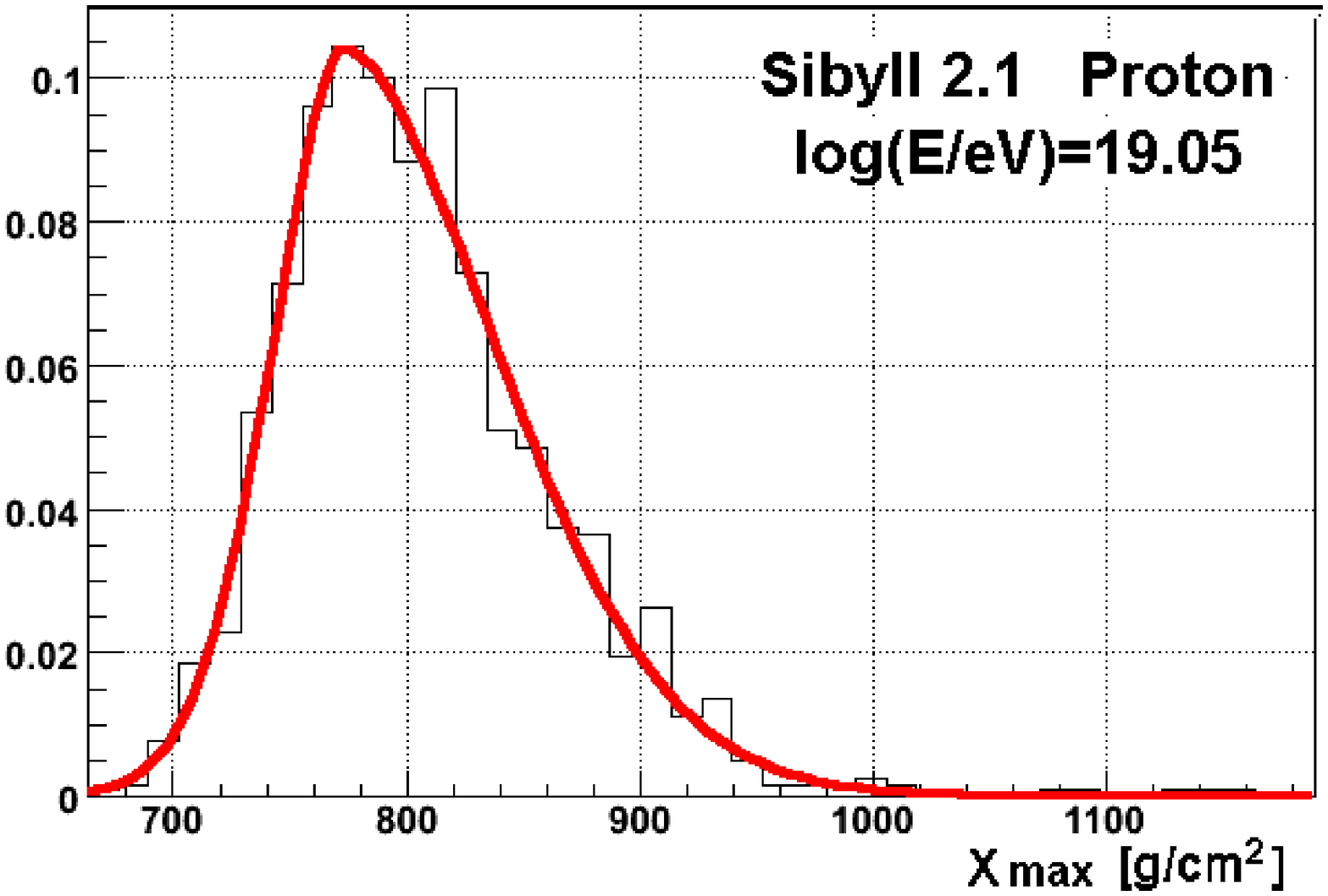}}
}
\caption{Examples of the fits with AGG function for the three parameters considered for different energy bins and hadronic 
interaction models.}
\label{fig:AGGFits}
\end{figure}

Fig. \ref{fig:AGGFits} shows one example of these fits for each parameter considered. As can be seen, the AGG function
allows us to fit asymmetric distributions with longer tails and sharper (or flatter) maximums compared to a Gaussian 
distribution. By using the distribution functions obtained in these fits, we could get the needed events to perform
the calculation with enough statistics.
Samples of events corresponding to a given primary type, parameter, energy bin and HIM are generated by sampling the
corresponding fitting function. The histograms $h_{p}$ and $h_{fe}$ used as the universe have 1000 events each. The
number of events of the test samples (used to calculate the error of the inferred composition) varies as a function
of primary energy because of the steepness of the spectrum. The number of events expected by Auger in 1 and 5 years 
of full operation are considered\footnote{For example, in one year of full operation and considering the spectrum 
reported in \cite{AugerSDSpectrumICRC07}, around 400 events are expected at $10^{19}$ eV and around 30 at 
$10^{19.6}$ eV}, and to reproduce real conditions, the number of events with available $X_{max}$ is $10\%$ of the 
total in the sample. The procedure to infer the composition is the same as explained before.
\begin{figure}[!bt]
\begin{center}
\subfigure{\includegraphics[width=10cm]{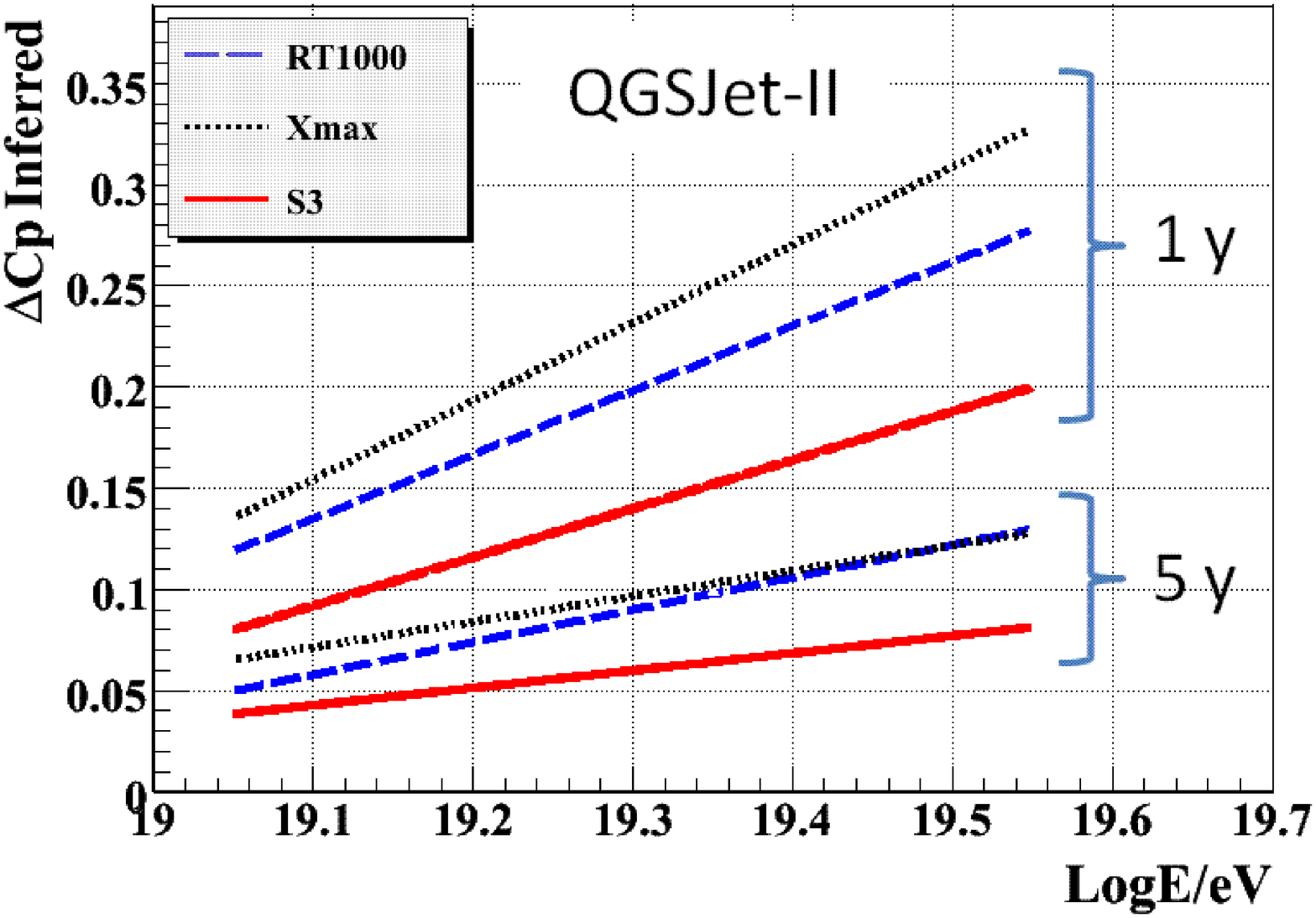} \label{fig:DeltaCp_QG}}
\subfigure{\includegraphics[width=10cm]{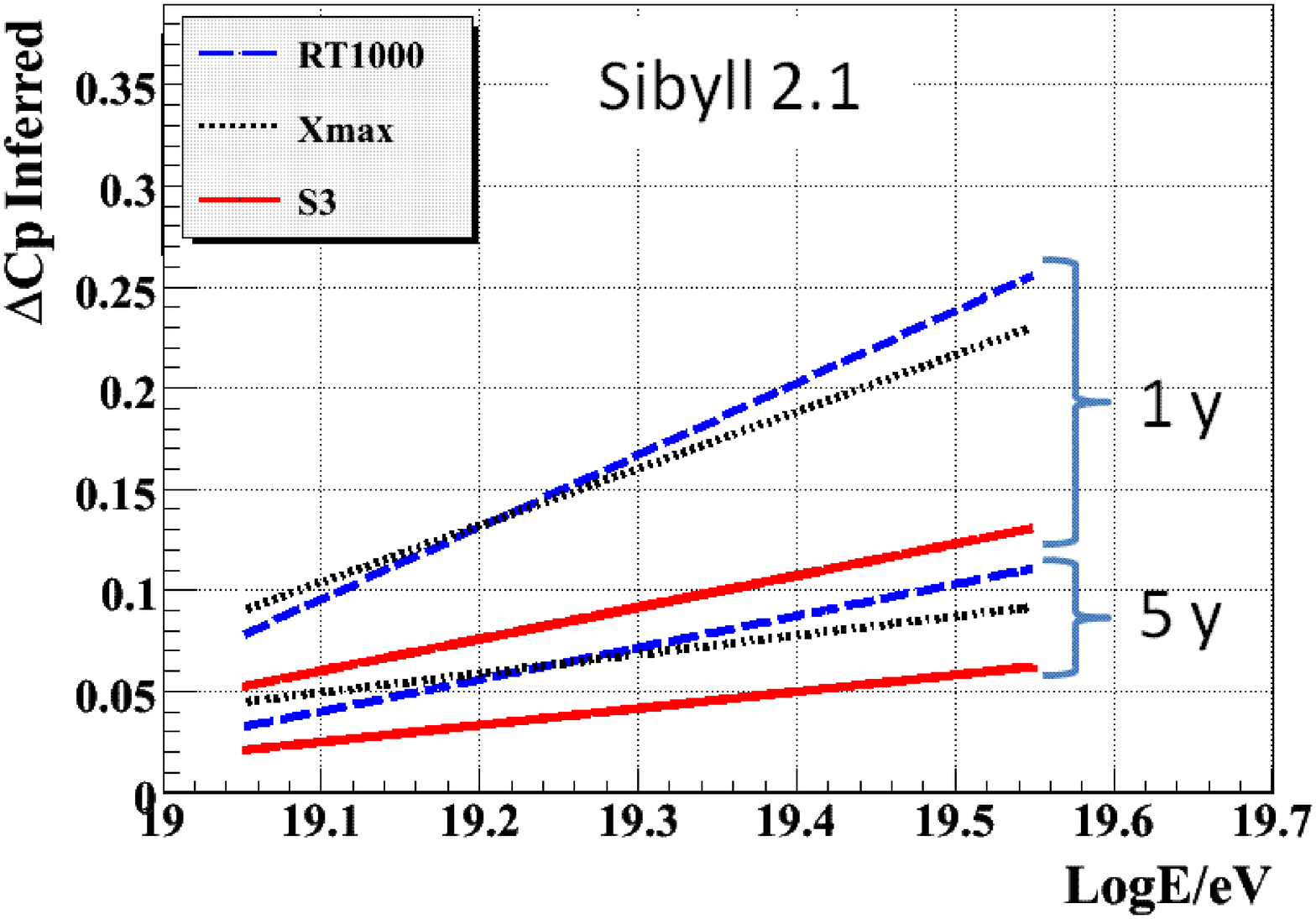} \label{fig:DeltaCp_Sib}}
\end{center}
\caption{Error in the inferred proton abundance determined by using $S_3$, $X_{max}$ and $t_{1/2}(r_0)$ for 1 and 5 years of Auger full operation time. Top: QGSJet-II. Down: Sibyll 2.1.
The statistics used for $X_{max}$ is only $10\%$ compared to that for surface parameters due to the limited duty cycle of the fluorescence telescopes.}
\label{fig:DeltaCp_AGG}
\end{figure}

As in previous case, there is no significant bias in the inferred proton abundance. Fig. \ref{fig:DeltaCp_AGG} 
shows the uncertainty (at $68\%$ of C.L.) on the determination of the proton abundance as a function of primary energy for samples of $C_p^{true} = 0.5$. 
It can be seen that the best results are obtained for $S_3$. As mentioned before, the errors corresponding to Sibyll 2.1 are smaller than for QGSJET-II because the shower to shower 
fluctuations are in general smaller for Sibyll 2.1.

As mentioned in Section \ref{Simulations}, the reconstructed energy considered in this work is obtained fluctuating the real one with a Gaussian uncertainty. This procedure does not 
take into account the possible correlation between the reconstructed energy and primary composition, which has the potential to degrade the discrimination power of any mass composition 
parameter. In order to assess the magnitude of such effect, we consider the energy estimator $S_{38}$ as in the case of Auger \cite{AugerHybridSpectrumICRC09}.

\begin{figure}[!bt]
\centerline{
\subfigure{\includegraphics[width=7cm]{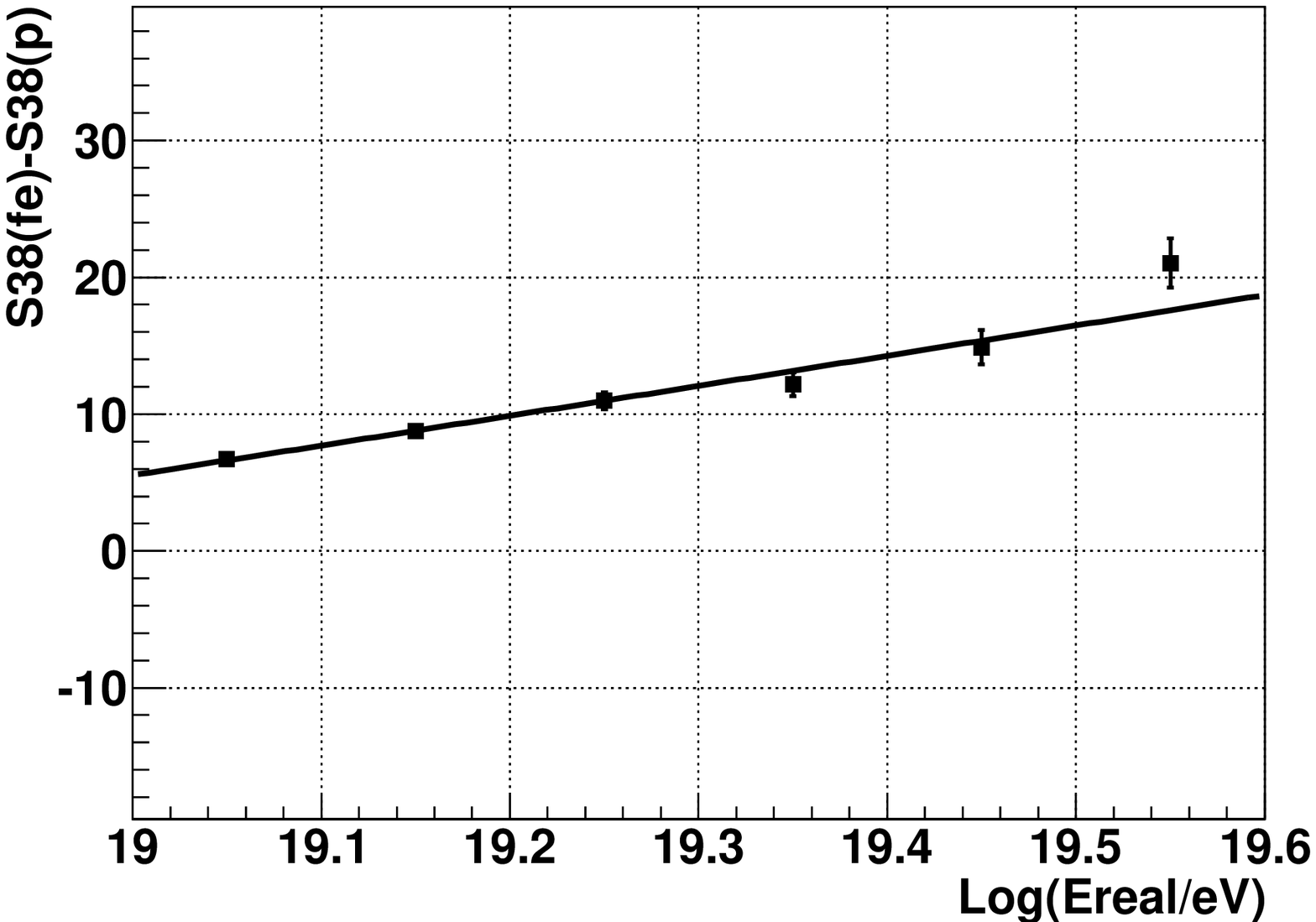} \label{fig:DeltaS38_vs_RealEnergy}}
\hfil
\subfigure{\includegraphics[width=7cm]{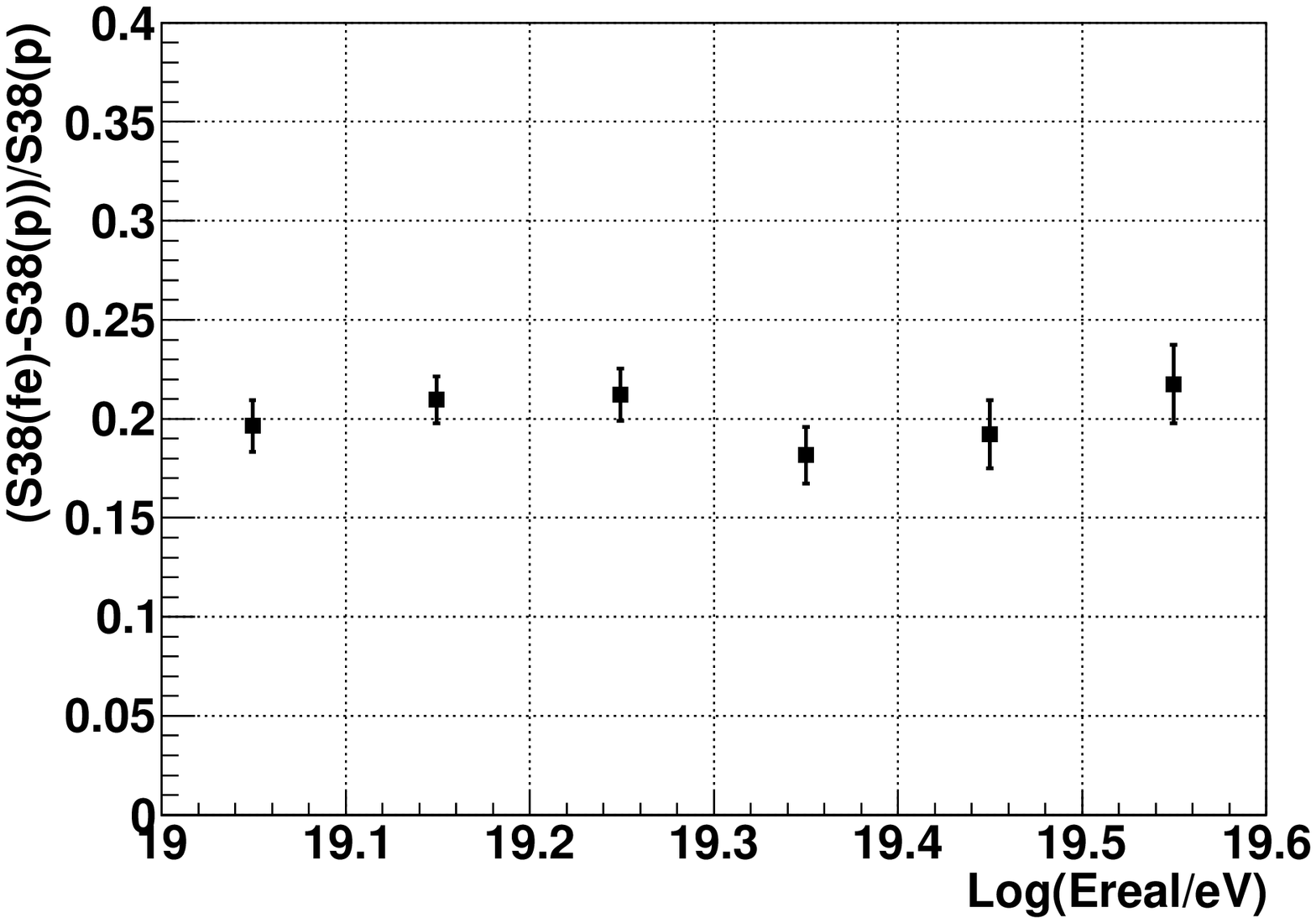} \label{fig:RelDeltaS38_vs_RealEnergy}}
}
\caption{$\Delta S_{38} = S_{38}(fe) - S_{38}(p)$ (left) and $\Delta S_{38} / S_{38}(p)$ (right) as a function of the real energy. The error bars are the $RMS/\sqrt N.$}
\end{figure}

It is expected that, for the same energy, different primaries will produce showers with different $S(r)$ parameter, giving rise to systematic effects on the reconstructed energy depending on the
unknown composition of the impinging cosmic ray flux. In fact, from the simulations, we calculate the difference in $S_{38}$ between iron and proton showers with the same real energy and,
as Fig. \ref{fig:DeltaS38_vs_RealEnergy} shows, $\Delta S_{38} = S_{38}(fe) - S_{38}(p)$ increases almost linearly with the real primary energy. The relative value $\Delta S_{38}/S_{38}(p)$
is also shown in Fig. \ref{fig:RelDeltaS38_vs_RealEnergy} (which is in agreement with \cite{Maris_Thesis}). This difference translates into a spurious difference, $\Delta E$, between the 
estimated energy for proton and Iron primaries of the same arrival energy. Therefore, in order to assess the influence of the correlation between the reconstructed energy and composition on 
the stability of our $S_3$ parameter, we modify the reconstructed energy of each iron event by adding to its originally estimated energy the systematic error, $\Delta E$, calculated from its own $S_{38}$. 
On the other hand, the procedure for the reconstruction of the energy of proton primaries is not modified. The energies thus obtained are used to repeat the process described above for composition 
discrimination in the case of surface parameters.

\begin{figure}[!bt]
\begin{center}
\includegraphics[width=10cm]{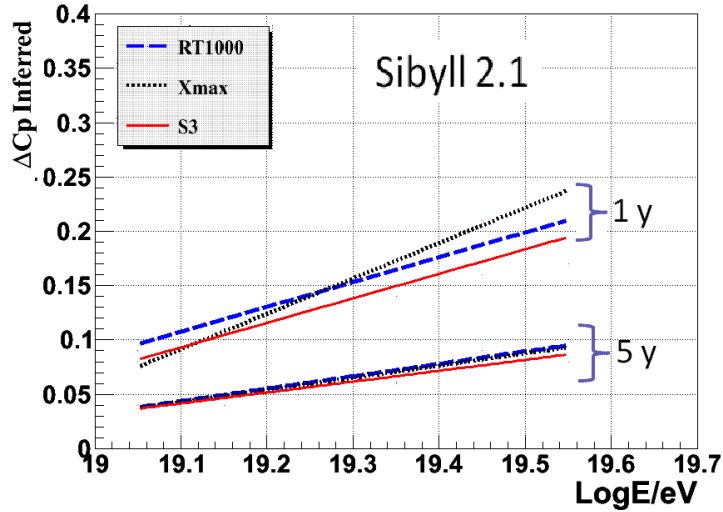}
\caption{As Fig. \ref{fig:DeltaCp_Sib} but considering the correlation between the reconstructed energy and the primary composition. See text for details.}
\label{fig:DeltaCp_Sib_Corr}
\end{center}
\end{figure}

Fig. \ref{fig:DeltaCp_Sib_Corr} shows the results obtained for $1$ and $5$ years of integration time. The error in the measured composition using $S_3$ increases but still remains smaller, 
or of the order of those obtained using rise time or $X_{max}$ as discriminator parameters. The results shown in Fig. \ref{fig:DeltaCp_Sib_Corr} were calculated using Sibyll 2.1 as 
hadronic interaction model, but are qualitatively similar for QGSJET-II.

\section{Conclusions}

A new family of parameters, that we call $S_b$, has been proposed for composition analysis in cosmic ray surface arrays. 
The parameters are determined on a shower-to-shower basis from the total signal deposited in each triggered detector and 
their distance to the shower axis. Despite the fact that surface composition parameters are usually more affected by systematic 
errors than the corresponding parameters obtained from the fluorescence technique, the former are of great interest because of their larger duty cycle (around 10 times larger).

$S_b$ is of general applicability: its discrimination power has been studied for different surface arrays, 
varying the distance between detectors and the array geometry (square and triangular grids). The separation
power of $S_b$ increases rapidly as the array spacing decreases and remains strong for any array geometry.

We have performed an extensive analytical study of the most relevant properties of $S_b$. 
We demonstrate that there is a well defined value of the parameter $b$ which maximizes the discrimination capability of $S_{b}$. In the case of a binary mixture or proton and Iron, 
that optimal value is $b \approx 3$. 

Furthermore, motivated by experimental results that suggest an excess of muon shower size with respect to numerical simulation expectations, 
we have analyzed the stability of $b$ under such conditions, showing that the optimal fit for $b$ is still $\sim 3$ and that, furthermore, 
the discrimination power of $S_{3}$, for an Fe-p mixture, is actually enhanced.

Numerical simulations have also been carried out to account for experimental conditions and the uncertainties involved in the reconstruction 
procedure. The previous analytical results have also been validated from the simulations which include the detector response and reconstruction 
efficiency. The numerical results support that $b \cong 3$ is the value that maximizes the discrimination power of $S_b$, in agreement with the 
analytical method, and independently of the array geometry or the distance between detectors.

Additionally, we demonstrate that $S_3$ is almost independent on zenith angle, which represents an advantage over other mass sensitive parameters, 
and it is almost linearly dependent on the primary energy.

A realistic comparison between $S_3$ and the most useful parameters from the surface and fluorescence technique, i.e., the rise time, 
$t_{1/2}$, and $X_{max}$ respectively, has been performed. Our simulations show that the accuracy on the determination of the composition 
obtained by using $S_{3}$ is, at constant integration time, greater than the one obtained using the other parameters considered. 
Even if a realistic correlation between the reconstructed energy and the primary composition is included in the analysis, the discrimination power 
of $S_{3}$, albeit degraded, still remains greater or of the order of the one obtained using the other parameters.

The application of $S_3$ or any other composition parameter in a real experiment is not straightforward. The most significant problems 
that could affect $S_3$ have been considered in this work, such as the energy uncertainty, zenith angle dependence, surface array density and 
geometry, the effect of detectors very far from the core and the uncertainties in the hadronic interaction models (specially in the muon 
component). On the other hand, $X_{max}$ is less prone to systematic errors but several circumstances must be considered to achieve reliable 
results. For instance, the longitudinal profile reconstruction suffers significant uncertainties coming from the fluorescence yield or possible 
direct Cherenkov light towards the telescopes. Atmospheric monitoring is also needed to account for the effect of clouds and aerosols. In hybrid 
experiments it requires an optimum synchronization with a surface detector as well. In addition, only those events whose $X_{max}$ is in the 
field of view of the telescopes could be properly reconstructed and this selection cut could introduce artificial biases that must be carefully 
corrected \cite{AugerXmax2010}.

\section{Acknowledgments}

All the authors have greatly benefited from their participation in the Pierre Auger Collaboration and its profitable 
scientific atmosphere. We want to thank the Offline team for providing Auger Offline packages tools used in this work. 
G. Ros thanks to Comunidad de Madrid for a F.P.I. fellowship, Universidad de Alcal\'a for a post-doctoral grant, and the 
ALFA-EC funds in the framework of the HELEN project. G. Ros also thanks to the Instituto de Ciencias Nucleares, UNAM, 
for its hospitality during several stays. Extensive numerical simulations were made possible by the use of the UNAM 
super-cluster \emph{Kanbalam}. G. Medina Tanco also acknowledges the support of the HELEN project. A. D. Supanitsky 
acknowledges a postdoctoral grant from UNAM. 

This work is partially supported by Spanish Ministerio de Educaci\'on y Ciencia under the projects 
FPA2006-12184-C02-02, FPA2008-04192-E, FPA2008-01698-E, FPA2009-11672 and CSD2007-00042,
Mexican PAPIIT-UNAM through grants IN115707 and IN115607 and CONACyT through grants 46999-F, 57772, CB-2007/83539.

\end{document}